\documentclass[fleqn,usenatbib,useAMS]{mnras}

\DeclareRobustCommand{\VAN}[3]{#2}
\let\VANthebibliography\thebibliography
\def\thebibliography{\DeclareRobustCommand{\VAN}[3]{##3}\VANthebibliography}

\usepackage{amsmath,amsfonts,amssymb} 
\usepackage{bm} 
\usepackage[table,  dvipsnames]{xcolor} 
\usepackage{hyperref} 
\usepackage{nicefrac} 
\usepackage{placeins}

\usepackage[varg]{txfonts} 
\usepackage[T1]{fontenc}
\usepackage[normalem]{ulem}

\hypersetup{
	pdftitle = {Planet-driven gap opening in ALMA disks: the role of dust dynamics},
	pdfauthor = {A. Ziampras, P. Sudarshan, C. P. Dullemond, V. Berta, M. Flock, R. P. Nelson, A. Mignone},
	pdfsubject = {},
	colorlinks = {true},
	linkcolor = [rgb]{0,0.35,0.7},
	citecolor = [rgb]{0,0.35,0.7},
	filecolor = [rgb]{0.61,0,0},
	urlcolor = [rgb]{0.61,0,0},
}

\usepackage{graphicx}  
\graphicspath{{.}}

\newcommand{\tensorGR}[1]{\overline{\bm{{#1}}}}
\newcommand{\DP}[2]{\frac{\partial{#1}}{\partial{#2}}}
\newcommand{\D}[2]{\frac{\text{d}{#1}}{\text{d}{#2}}}

\newcommand{\G}{\text{G}}
\newcommand{\Mstar}{M_\star}
\newcommand{\Lstar}{L_\star}
\newcommand{\Rp}{R_\mathrm{p}}
\newcommand{\Mp}{M_\mathrm{p}}

\newcommand{\Tb}{T_\mathrm{b}}
\newcommand{\Mth}{M_\mathrm{th}}
\newcommand{\Msun}{\mathrm{M}_\odot}
\newcommand{\Lsun}{\mathrm{L}_\odot}
\newcommand{\Mjup}{\mathrm{M}_\mathrm{J}}
\newcommand{\Rgas}{\mathcal{R}}
\newcommand{\cs}{c_\mathrm{s}}

\newcommand{\OmegaK}{\Omega_\mathrm{K}}
\newcommand{\vK}{u_\mathrm{K}}

\newcommand{\tauR}{\tau_\mathrm{R}}
\newcommand{\tauP}{\tau_\mathrm{P}}

\newcommand{\taueff}{\tau_\mathrm{eff}}
\newcommand{\kappaR}{\kappa_\mathrm{R}}
\newcommand{\kappaP}{\kappa_\mathrm{P}}

\newcommand{\rhomid}{\rho_\mathrm{mid}}
\newcommand{\sigmaSB}{\sigma_\mathrm{SB}}
\newcommand{\vel}{\bm{u}}

\newcommand{\tcool}{t_\mathrm{cool}}

\newcommand{\Qcool}{Q_\mathrm{cool}}

\newcommand{\Qrad}{Q_\mathrm{rad}}

\newcommand{\Sigmag}{\Sigma_\mathrm{g}}
\newcommand{\Sigmad}{\Sigma_\mathrm{d}}
\newcommand{\velg}{\vel_\mathrm{g}}
\newcommand{\veld}{\vel_\mathrm{d}}

\newcommand{\St}{\mathrm{St}}
\newcommand{\ts}{t_\mathrm{s}}

\newcommand{\ad}{a_\mathrm{d}}
\newcommand{\brhod}{\bar{\rho}_\mathrm{d}}

\newcommand{\pluto}{\texttt{PLUTO}}
\newcommand{\dw}{\texttt{DUSTYWAVE}}

\title[Dusty substructures by planets]{Dusty substructures induced by planets in ALMA disks:\\how dust growth and dynamics changes the picture}

\author[A.~Ziampras et al.]{Alexandros~Ziampras$^{1,2}$\thanks{E-mail: \texttt{a.ziampras@qmul.ac.uk}},
Prakruti~Sudarshan$^{3}$\thanks{Correspondence for the dust fluid module: \texttt{sudarshan@mpia.de}},
Cornelis~P.~Dullemond$^{4}$,
Mario~Flock$^{3}$,
Vittoria~Berta$^{5}$,
\newauthor
Richard~P.~Nelson$^{1}$,
Andrea~Mignone$^{5}$
\\
$^{1}$Astronomy Unit, School of Physics and Astronomy, Queen Mary University of London, London E1 4NS, UK\\
$^{2}$Ludwig-Maximilians-Universit{\"a}t M{\"u}nchen, Universit{\"a}ts-Sternwarte, Scheinerstr.~1, 81679 M{\"u}nchen, Germany\\
$^{3}$Max-Planck-Institut f{\"u}r Astronomie, K{\"o}nigstuhl 17, 69117 Heidelberg, Germany\\
$^{4}$Institute for Theoretical Astrophysics, Center for Astronomy (ZAH), Heidelberg University, Albert Ueberle Str.~2, 69120 Heidelberg,
Germany\\
$^{5}$Dipartimento di Fisica, Universit{\`a} di Torino, via Pietro Giuria 1, 10125 Torino, Italy\\
}

\date{Accepted XXX. Received YYY; in original form ZZZ}
\pubyear{2023}

\begin{document}
\label{firstpage}
\pagerange{\pageref{firstpage}--\pageref{lastpage}}
\maketitle

\begin{abstract}
	Protoplanetary disks exhibit a rich variety of substructure in millimeter continuum emission, often attributed to unseen planets. As these planets carve gaps in the gas, dust particles can accumulate in the resulting pressure bumps, forming bright features in the dust continuum. We investigate the role of dust dynamics in the gap-opening process with 2D radiation hydrodynamics simulations of planet--disk interaction and a two-population dust component modeled as a pressureless fluid. We consider the opacity feedback and backreaction due to drag forces as mm grains accumulate in pressure bumps at different stages of dust growth. We find that dust dynamics can significantly affect the resulting substructure driven by the quasi-thermal-mass planet with $\Mp/\Mstar=10^{-4}$. Opacity feedback causes nonaxisymmetric features to become more compact in azimuth, whereas the drag-induced backreaction tends to dissolve nonaxisymmetries. For our fiducial model, this results in multiple concentric rings of dust rather than the expected vortices and corotating dust clumps found in models without dust feedback. A higher coagulation fraction disproportionately enhances the effect of dust opacity feedback, favoring the formation of crescents rather than rings. Our results suggest that turbulent diffusion is not always necessary to explain the rarity of observed nonaxisymmetric features, and that incorporating dust dynamics is vital for interpreting the observed substructure in protoplanetary disks. We also describe and test the implementation of the publicly-available dust fluid module in the \pluto{} code.
\end{abstract}



\section{Introduction}
\label{sec:introduction}

The ALMA large program DSHARP \citep{andrews-etal-2018} has provided a wealth of high-resolution images of protoplanetary disks, revealing a rich variety of structures in millimeter emission. One of the most striking features is the ubiquitous presence of gaps and rings in the dust distribution \citep{huang-etal-2018}, which can be attributed to a variety of mechanisms such as dust trapping around snowlines \citep{owen-2020}, excavation of gas due to magnetic effects \citep{bethune-etal-2017,riols-etal-2020}, or the presence of unseen planets \citep[e.g.,][]{zhang-etal-2018}. The latter mechanism is particularly interesting given the direct observation of two planets in the disk around the system PDS~70 \citep{keppler-etal-2018,haffert-etal-2019}, the indirect evidence for planets in other disks through gas kinematics \citep[e.g.,][]{teague-etal-2018}, and the theoretical predictions of planet formation models \citep[e.g.,][]{zhang-etal-2018,bae-etal-2019}.

Planets interact gravitationally with their natal disks, exciting characteristic spiral density waves in the gas \citep{goldreich-tremaine-1979,ogilvie-lubow-2002}. These waves steepen into shocks as they propagate, depositing angular momentum into the disk and leading to the formation of gaps in the gas distribution for a sufficiently massive planet \citep{rafikov-2002}. The resulting pressure maxima in the gas can trap dust particles, forming bright rings in dust continuum emission \citep{pinilla-etal-2012}. The planetary mass required to open a gap in the gas  \citep[``thermal mass'', see][]{rafikov-2002} is typically of the order of a sizable fraction of a Jupiter, heavily dependent on the disk properties \citep{crida-etal-2006} but nevertheless small enough to make a direct probe of the planet's presence difficult. Thus, understanding the process of planet-driven gap opening is crucial for interpreting the observed structures in protoplanetary disks.

Recent work has shown that, while gap opening is a robust process for massive enough planets \citep[e.g.,][]{rafikov-2002,crida-etal-2006}, it is sensitive to the disk's thermo- and hydrodynamics. The importance of radiative cooling has been highlighted by several studies \citep{miranda-rafikov-2019,miranda-rafikov-2020a,miranda-rafikov-2020b,ziampras-etal-2020b,ziampras-etal-2023a,zhang-zhu-2020,zhang-etal-2024}, which have shown that for intermediate cooling timescales ($\tcool\sim\OmegaK^{-1}$) radiative cooling can suppress the formation of secondary gaps, instead leading to the formation of a single, deep gap. For massive enough disks that the disk self-gravity becomes important \citep[$Q\gtrsim2$, see][]{toomre-1964}, planet-driven spirals carry a higher angular momentum flux and gap opening is enhanced \citep{zhang-zhu-2020}. Furthermore, magnetic fields can enhance gap opening due to both the accumulation of magnetic pressure in the gap and the erosion of the gap edge by magnetic winds \citep[e.g.,][]{wafflard-fernandez-lesur-2023,aoyama-bai-2023}. In this work, we will focus on the effects of radiative cooling in the non-self-gravitating, magnetically-inactive limit.

In the aforementioned studies on cooling effects, the role of dust has been often reduced to a carrier of opacity for small, sub-micron grains, assuming a constant dust-to-gas ratio, and a tracer of substructure for large, millimeter- to centimeter-sized grains \citep[e.g.,][]{zhang-etal-2018,bae-etal-2019}. Nevertheless, dust growth models predict that dust particles can quickly grow to sizes of a few millimeters in the outer disk \citep{birnstiel-etal-2012,stammler-birnstiel-2022,birnstiel-2023,dominik-dullemond-2024}, depleting the fraction of small grains and reducing the overall dust opacity. In itself, this can have a significant impact on the gap opening process by enabling the gas to cool more efficiently.

At the same time, however, the accumulation of large grains in a ring near the gap edge can lead to a significant local enhancement of the dust mass. This has implications on the total dust opacity \citep[see][]{binkert-etal-2021,binkert-etal-2023,szulagyi-etal-2022,krapp-etal-2024}, and especially so for lower temperatures where small grains contribute less to the opacity \citep[$\kappa_\text{small}\propto T^{1.5\text{--}2}$, see e.g.,][]{semenov-etal-2003,birnstiel-etal-2018}. The higher local dust-to-gas ratio in large grains can also drive significant momentum exchange between gas and dust \citep[e.g.,][]{weber-etal-2018,weber-etal-2019,hammer-lin-2023}. The combination of these effects can lead to a complex dynamical interplay between the dust and gas dynamics, which could be important in the context of planet-driven gap opening.

In this work, we aim to investigate the role of dust dynamics in the process of planet-driven gap opening in protoplanetary disks. With numerical simulations of planet--disk interaction using radiation hydrodynamics, we will explore the effects of dust growth and gas--dust interaction on the gap-opening process, with a focus on the outer, well-resolved regions of the disk ($\sim$$50\text{--}100$\,au).

The paper is structured as follows. In Sect.~\ref{sec:physics-numerics} we describe the physics and numerics of our models, and in Sect.~\ref{sec:dust-opacity} we discuss our dust opacity model. In Sect.~\ref{sec:results}, we present our results, and in Sect.~\ref{sec:discussion} we discuss their implications. We summarize our findings in Sect.~\ref{sec:summary}.

\section{Physics and numerics}
\label{sec:physics-numerics}

In this section we describe the physical framework and numerical setup of our models. We also discuss the initial conditions and parameters of our simulations.

\subsection{Radiation hydrodynamics}

We consider a vertically-integrated distribution of perfect gas with mean molecular weight $\mu=2.35$, adiabatic index $\gamma=7/5$, surface density $\Sigmag$, vertically-integrated pressure $P$ and velocity field $\velg$ in a Keplerian disk around a star with mass $\Mstar$ and luminosity $\Lstar$. The Navier--Stokes equations of hydrodynamics read \citep{tassoul-1978}
\begin{subequations}
	\label{eq:navier-stokes}
	\begin{align}
		\label{eq:navier-stokes-1}
		\D{\Sigmag}{t} = -\Sigmag\nabla\cdot\velg,
	\end{align}
	\begin{align}
	\label{eq:navier-stokes-2}
		\Sigmag\D{\velg}{t} =-\nabla P -\Sigmag\nabla\Phi +\nabla\cdot\bm{\tensorGR{\sigma}},\qquad \Phi=\Phi_\star+\Phi_\mathrm{p},
	\end{align}
	\begin{align}
	\label{eq:navier-stokes-3}
		\D{e}{t}=-\gamma e\nabla\cdot\velg+Q_\mathrm{visc}+Q_\mathrm{irr} + Q_\mathrm{cool} + Q_\mathrm{rad},
	\end{align}
\end{subequations}
where $\text{d}/\text{d}t=\partial/\partial t + \vel\cdot\nabla$ is the material derivative, $e = P/(\gamma-1)$ is the internal energy density, $\bm{\tensorGR{\sigma}}$ is the viscous stress tensor, $\Phi_\star = -\G\Mstar/R$ is the gravitational potential of the star at distance $R$, and $\G$ is the gravitational constant. Through the above we can define the isothermal sound speed $\cs = \sqrt{P/\Sigmag}$ and the pressure scale height $H = \cs/\OmegaK$, where $\OmegaK = \sqrt{\G\Mstar/R^3}$ is the Keplerian angular velocity. The temperature is given by $T=\mu\cs^2/\Rgas$, where $\Rgas$ is the ideal gas constant, and the disk aspect ratio is $h=H/R$.

The source terms $Q_\mathrm{visc}$, $Q_\mathrm{irr}$, $Q_\mathrm{cool}$ and $Q_\mathrm{rad}$ represent viscous dissipation, heating due to stellar irradiation, radiative cooling through the disk surfaces, and radiative diffusion across the disk plane, respectively:
\begin{subequations}
	\label{eq:source-terms}
	\begin{align}
		\label{eq:source-terms-1}
		Q_\mathrm{visc} = \frac{1}{2\nu\Sigmag}\mathrm{Tr}(\tensorGR{\sigma}^2) \approx \frac{9}{4}\nu\Sigmag\OmegaK^2,\quad \nu=\alpha\sqrt{\gamma}\cs H,
	\end{align}
	\begin{align}
		\label{eq:source-terms-2}
		Q_\mathrm{irr} = 2\frac{\Lstar}{4\pi R^2} (1-\epsilon)\frac{\theta}{\taueff}, \quad \theta = R\D{h}{R}\approx \frac{2h}{7},
	\end{align}
	\begin{align}
		\label{eq:source-terms-3}
		Q_\mathrm{cool} = -2\frac{\sigmaSB T^4}{\taueff}, \quad \taueff = \frac{3\tauR}{8} + \frac{\sqrt{3}}{4} + \frac{1}{4\tauP}, \quad \tau_i = \frac{\kappa_i\Sigmag}{2},
	\end{align}
	\begin{align}
		\label{eq:source-terms-4}
		Q_\mathrm{rad} = \sqrt{2\pi}H\nabla\cdot \left(\lambda\frac{4\sigmaSB}{\kappaR\rhomid}\nabla T^4\right), \quad \rhomid = \frac{1}{\sqrt{2\pi}}\frac{\Sigmag}{H}.
	\end{align}
\end{subequations}
Here, we have adopted the $\alpha$ prescription for the kinematic viscosity $\nu$ \citep{shakura-sunyaev-1973}, the irradiation model by \citet{menou-goodman-2004}, the effective optical depth $\taueff$ by \citet{hubeny-1990}, and the flux-limited diffusion approximation for the radiative diffusion term \citep[FLD,][]{levermore-pomraning-1981} with the flux limiter $\lambda$ given by \citet{kley-1989}. In the above equations, $\sigmaSB$ is the Stefan--Boltzmann constant, $\epsilon=1/2$ is the disk albedo, $\theta$ the flaring angle, $\rhomid$ the volume density at the midplane, and $\kappaR$, $\kappaP$ are the Rosseland and Planck mean opacities. For a detailed discussion of the radiative terms, we refer the reader to \citet{ziampras-etal-2023a}.

The planet's gravitational potential $\Phi_\mathrm{p}$ is given by a Plummer potential
\begin{equation}
	\Phi_\mathrm{p} = -\frac{\G\Mp}{\sqrt{d^2+\epsilon^2}}, \quad \bm{d} = \bm{R} - \bm{R}_\text{p},
\end{equation}
where $\Mp$ is the planet mass and $\epsilon=0.6H$ is the smoothing length, chosen to account for the vertical stratification of the disk \citep{mueller-etal-2012}. The planet is assumed to be on a fixed circular orbit and does not accrete material. We include the indirect term due to the planet--star system orbiting their common center of mass, but not the acceleration of the star due to nonaxisymmetric features in the disk.

\subsection{Dust as a pressureless fluid}
\label{sub:dust-model}

We include the dust component as a pressureless fluid with surface density $\Sigmad$ and velocity field $\veld$. The dust couples to the gas through aerodynamic drag, which is modeled as a friction force between the two fluid components. The equations that dictate the dynamics for dust particles of size $\ad$ and bulk density $\brhod$ are given by
\begin{subequations}
	\label{eq:dust-evolution}
	\begin{align}
		\label{eq:dust-evolution-1}
		\D{\Sigmad}{t} = -\Sigmad\nabla\cdot\veld,
	\end{align}
	\begin{align}
		\label{eq:dust-evolution-2}
		\Sigmad\D{\veld}{t} = - \Sigmad\nabla\Phi - \Sigmad\frac{\veld-\velg}{\St}\OmegaK,
	\end{align}
\end{subequations}
where $\St$ is the Stokes number or dimensionless stopping time, defined in the Epstein regime as \citep[e.g.,][]{armitage-2009}
\begin{equation}
	\label{eq:stokes-number}
	\St = \frac{\pi}{2}\frac{\ad\brhod}{\Sigmag}.
\end{equation}
To account for the dust feedback on the gas, we include the dust backreaction term in Eq.~\eqref{eq:navier-stokes-2} as
\begin{equation}
	\label{eq:back-reaction}
	\Sigmag\DP{\velg}{t} = - \Sigmad\frac{\velg-\veld}{\St}\OmegaK.
\end{equation}

\subsection{Numerical setup}
\label{sub:numerics}

We use the \pluto{} \texttt{v.4.4} code \citep{mignone-etal-2007} to solve the equations of radiation hydrodynamics including a dust component on a 2D polar ($R,\varphi$) grid. The grid extends radially with logarithmic spacing between $R\in[0.1,4]$\,$R_0$ and covers the full azimuthal extent $\varphi\in[0,2\pi]$, with $N_R\times N_\varphi=672\times1152$ cells. For our setup this achieves a resolution of 10 cells per $H$ at $R_0$ in both directions, which is sufficient for our study. We also make use of the FARGO algorithm \citep{masset-2000,mignone-etal-2012}, which subtracts the background rotation before calculations to greatly relax timestep requirements. We use the HLLC Riemann solver \citep{toro-etal-1994} with the limiter by \citet{vanleer-1974}, and a second-order Runge--Kutta time integration scheme. The FLD term in Eq.~\eqref{eq:source-terms-4} is implemented following \citet{ziampras-etal-2020a}, and dust is modeled as a pressureless fluid (see Sec.~\ref{sub:dust-model}) with an implicit first-order accurate solver. The implementation and several verification tests for the dust module are presented in Appendix~\ref{apdx:dust-pluto}.

We model our disk after the HD~163296 system, with $\Mstar=2.04\,\Msun$, $\Lstar=16.98\,\Lsun$ \citep{andrews-etal-2018}, and $R_0=48$\,au, with $\Sigma(R_0)=10$\,g/cm${}^2$ \citep{zhang-etal-2018}. A balance among the cooling, viscous, and irradiation heating terms then sets the disk aspect ratio to $h\approx0.057\,(R/R_0)^{2/7}$ \citep{chiang-goldreich-1997}, for a midplane temperature of $T(R_0)\approx35$\,K. We place a planet of mass $\Mp=0.2\,\Mjup = 10^{-4}\Mstar$ at $R_0$, and allow it to grow to its full mass over 100~orbits using the formula by \citet{devalborro-etal-2006}. This planet mass corresponds to approximately $0.8\,\Mth$, where $\Mth=\nicefrac{2}{3}h^3\Mstar$ is the thermal mass \citep{rafikov-2002}.
As we are not interested in viscous smoothing effects, we use a vanishingly small $\alpha=10^{-6}$ and do not consider dust diffusion. We nevertheless make a note on the implications of higher $\alpha=10^{-5}$--$10^{-3}$ in Appendix~\ref{apdx:alpha}.

Our initial conditions for the gas are
\begin{equation}
	\label{eq:initial-conditions}
	\Sigma_\text{g,0} = 10\,\text{g}/\text{cm}^2\,\left(\frac{R}{R_0}\right)^{-1},\quad T_0 = 34.64\,K\,\left(\frac{R}{R_0}\right)^{-3/7}
\end{equation}
with a Keplerian velocity field, corrected for gas pressure support. As discussed in Section~\ref{sec:dust-opacity}, we adopt a two-population dust model with ``small'' and ``big'' grains $\ad^\text{small}=0.1\,\mu$m and $\ad^\text{big}=1$\,mm, with $\St_0^\text{small}\approx3.3\times10^{-6}$ and $\St_0^\text{big}\approx3.3\times10^{-3}$ at $R_0$. Since we can only model one dust fluid, we assume that the small grains are perfectly coupled to the gas and their surface density is always given by $\Sigmad^\text{small}=(1-X_0)\varepsilon_0\Sigmag$, where $X=\Sigmad^\text{big}/(\Sigmad^\text{small}+\Sigmad^\text{big})$ is a ``coagulation fraction'' and $\varepsilon_0=\Sigmad/\Sigmag=0.01$ is the initial dust-to-gas ratio. The big grains are initialized with $\Sigma_\text{d,0}^\text{big}=X_0\varepsilon_0\Sigma_\text{g,0}$, and evolved according to Eq.~\eqref{eq:dust-evolution}.

We note that $X$, being expressed as the dust mass fraction in big grains, is subject to the radial drift of large grains during the simulation. In that sense, while $X_0$ represents a coagulation fraction at $t=0$ by setting the initial dust mass fraction in big grains, the time-dependent value of $X$ is no longer tied to coagulation and will evolve as big grains drift radially or accumulate in pressure bumps.

The domain is periodic in the $\varphi$ direction, and we enforce the initial conditions for the gas density and velocity at the inner and outer radial boundaries. We employ wave-killing zones radially between $R\in[0.1,0.125]\cup[3.4,4]\,R_0$ with a damping timescale of 0.1 boundary orbits \citep{devalborro-etal-2006}. We choose to not damp or reset the dust surface density at the outer boundary, and instead set an outflow boundary condition there, to avoid creating an infinite dust mass reservoir. We run our simulations for 1100 orbits, which is sufficient for the planet to open a gap in the gas and for the dust to settle into rings around the planet. This corresponds to a physical time of $\sim$0.26\,Myr. Our choice of simulation time is motivated by the time required to drain the dust reservoir in the outer disk, but we note that the features discussed here (rings, vortices) are already established within $\sim$400 orbits.

Since our aim is to investigate the effect of dust dynamics on the gap-opening process, we carry out the following families of models:
\begin{itemize}
	\item In our \emph{fiducial} models we include all radiative terms in the energy equation Eq.~\eqref{eq:navier-stokes-3}. The dust opacity is computed as $\kappa=(1-X_0)\kappa^\text{small} + X_0\kappa^\text{big}$, assuming that the redistribution of big grains does not affect the opacity.
	\item In models with \emph{opacity feedback} we recalculate $X=\Sigmad^\text{big}/(\Sigmad^\text{small}+\Sigmad^\text{big})$ at each timestep and update the dust opacity accordingly.
	\item In models with \emph{backreaction} we further include the term in Eq.~\eqref{eq:back-reaction}. These correspond to our most realistic models.
	\item As a control, we also run models with a \emph{locally isothermal} equation of state, where we set $T=T_0$ instead of evolving Eq.~\eqref{eq:navier-stokes-3}.
\end{itemize}

To explore the effects of grain growth, opacity feedback, and gas--dust interaction, we therefore carry out a set of 10 models, listed in Table~\ref{table:models}. We will present the results of these simulations in Sect.~\ref{sec:results}.

\begin{table}
	\centering
	\caption{List of models in our study. Radiative models include the energy equation in Eq.~\eqref{eq:navier-stokes-3} with all radiative terms, while in locally isothermal models we instead set $T=T_0$. The value of the coagulation fraction $X_0$ is varied to explore the effects of dust growth. For all models we set $\alpha=10^{-6}$.}
	\label{table:models}
	\begin{tabular}{lcccc}
		\hline
		Model & EOS & $X_0$ & $\kappa$ feedback & backreaction\\
		\hline
		\texttt{iso} & loc.~isothermal & 0.9 & N/A & No\\
		\texttt{iso-b} & loc.~isothermal & 0.9 & N/A & Yes\\
		\texttt{rad} & radiative & 0.9 & No & No\\
		\texttt{rad-o} & radiative & 0.9 & Yes & No\\
		\texttt{rad-ob} & radiative & 0.9 & Yes & Yes\\
		\texttt{rad-X0.99} & radiative & 0.99 & No & No\\
		\texttt{rad-o-X0.99} & radiative & 0.99 & Yes & No\\
		\texttt{rad-ob-X0.99} & radiative & 0.99 & Yes & Yes\\
		\texttt{rad-X0.01} & radiative & 0.01 & No & No\\
		\texttt{rad-o-X0.01} & radiative & 0.01 & Yes & No\\
		\texttt{rad-ob-X0.01} & radiative & 0.01 & Yes & Yes\\
		\hline
	\end{tabular}
\end{table}

\section{Dust opacity model}
\label{sec:dust-opacity}

We model the dust distribution as a two-population system, with small grains of size $\ad^\text{small}=0.1\,\mu$m and large grains of size $\ad^\text{big}=1\,\text{mm}$. To compute the dust opacity we use the code \texttt{OpTool} \citep{dominik-etal-2021}, using the Distribution of Hollow Spheres approach \citep[DHS,][]{min-etal-2005}. We assume that dust grains are composed of a mixture of 87\% amorphous pyroxenes and 13\% amorphous carbon \citep{zubko-etal-1996} and have a porosity of 25\%, for a resulting bulk density of $\brhod=2.08\,\text{g/cm}^3$.

\begin{figure}
	\centering
	\includegraphics[width=\columnwidth]{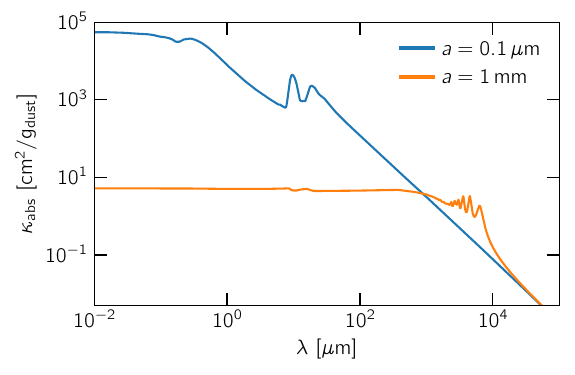}
	\caption{Dust absorption opacity as a function of wavelength for ``small" (sub-micron, blue line) and ``big" (millimeter, orange line) grains, as generated by \texttt{OpTool} for our choice of dust composition.}
	\label{fig:opacity-wavelength}
\end{figure}

\begin{figure}
	\centering
	\includegraphics[width=\columnwidth]{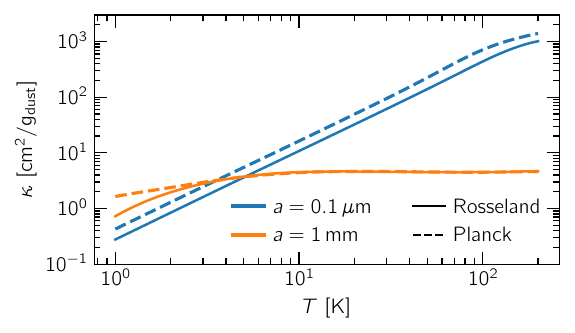}
	\caption{Rosseland (solid lines) and Planck (dashed) mean opacities as a function of temperature for small (blue) and big (orange) grains. Small grains follow a power law $\propto T^{1.6}$, while big grains have a roughly constant opacity.}
	\label{fig:opacity-temperature}
\end{figure}

The absorption opacity $\kappa_\nu$ for both grains as a function of wavelength is shown in Fig.~\ref{fig:opacity-wavelength}. We then compute the Rosseland and Planck mean opacities as a function of temperature for each species
\begin{equation}
	\label{eq:opacity-mean}
	\kappa_\text{P}^i = \frac{\int_0^\infty \kappa_\nu^i B(\nu,T)\,\text{d}\nu}{\int_0^\infty B(\nu,T)\,\text{d}\nu},\quad \kappa_\text{R}^i = \frac{\int_0^\infty u(\nu,T)\,\text{d}\nu}{\int_0^\infty (\kappa_\nu^i)^{-1} u(\nu,T)\,\text{d}\nu}, \quad u = \DP{B}{T},
\end{equation}
where $B(\nu,T)$ is the Planck black-body radiance. The resulting mean opacities are shown in Fig.~\ref{fig:opacity-temperature}. Conveniently, all mean opacities here are approximated well by power laws in temperature, with
\begin{align}
\label{eq:opacity-power-law}
&\kappaR^\text{small} \approx 0.27~T_\text{K}^{1.6}\,\text{cm}^2/\text{g},\quad \kappaP^\text{small} \approx 0.41~T_\text{K}^{1.6}\,\text{cm}^2/\text{g},\\
&\kappaR^\text{big} \approx \kappaP^\text{big}\approx 4.6\,\text{cm}^2/\text{g},
\end{align}
where $T_\text{K}$ is the temperature in Kelvin. 
In our models we will use the opacity power laws given above to compute the dust opacity of each species, and then calculate the total opacity per gram of dust as
\begin{equation}
	\label{eq:avg-kappa}
	\kappa_\text{R,P} = \frac{\kappa_\text{R,P}^\text{small}\Sigmad^\text{small} + \kappa_\text{R,P}^\text{big}\Sigmad^\text{big}}{\Sigmad^\text{small} + \Sigmad^\text{big}} = (1-X)\,\kappa_\text{R,P}^\text{small} + X\,\kappa_\text{R,P}^\text{big},
\end{equation}
where we have defined the fraction of dust mass in big grains $X=\Sigmad^\text{big}/(\Sigmad^\text{small}+\Sigmad^\text{big})$ as in Sect.~\ref{sub:numerics}. This approach is similar to that of \citet{binkert-etal-2023} and \citet{krapp-etal-2024}.

\begin{figure}
	\centering
	\includegraphics[width=\columnwidth]{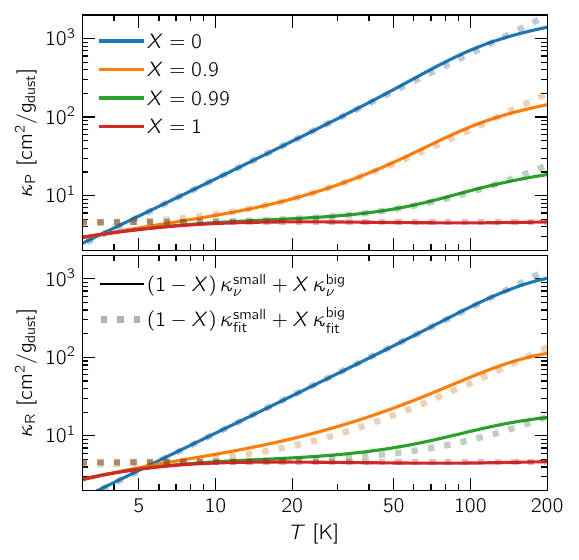}
	\caption{Approximation of the Rosseland and Planck mean opacities as a function of temperature for a two-population dust model with different coagulation fractions $X$, ranging from 0 (small grains only) to 1 (big grains only). The approximation works well for $T\gtrsim5$\,K, which is well within the scope of this work ($T\sim20$--100\,K)}.
	\label{fig:opacity-approximation}
\end{figure}

In principle, this approach is valid only in the optically thin limit (i.e., for the Planck mean opacity), while computing an equivalent Rosseland mean is not as straightforward. Nevertheless, we find that Eq.~\eqref{eq:avg-kappa} approximates the opacity of the two-population model well, as shown in Fig.~\ref{fig:opacity-approximation}. In this figure we compare the exact Rosseland and Planck means (obtained by averaging the density-weighted sum of absorption opacities $\kappa_\nu$ over frequency) to the approximation given by Eq.~\eqref{eq:avg-kappa} (obtained by the density-weighted sum of frequency-averaged opacities) for different coagulation fractions $X$. 

We note that the ratio of opacities between small and big grains is $\kappaR^\text{small}/\kappaR^\text{big}\sim 10$ for $T\sim 30$\,K. This implies that, assuming that 90\% of the sub-micron grains have coagulated into mm-sized ``pebbles'' (or $X=0.9$), the two populations would contribute equally to the opacity reservoir, with the big grains supplying most of the opacity if $X=0.99$ instead. Such levels of depletion of small grains are typical in dust growth models \citep{birnstiel-2023}. This can lead to a dynamic change of the dust opacity in the disk due to the redistribution of large grains around pressure bumps, which becomes more important the more the disk is depleted of small grains and at lower temperatures.

Finally, our dust model does not consider changes in the dust composition due to the sublimation of different volatiles around snowlines. We therefore limit our analysis to $T\lesssim150$\,K, as the opacity of the dust grains is expected to change at around the water snowline at $\sim$170\,K \citep[e.g.,][]{lin-papaloizou-1985,bell-lin-1994}. This is not a problem for our models, as we are interested in the cold outer disk regions (for our simulation domain, we have $T\sim20$--100\,K). Nevertheless, we found that including a water ice mantle with a mass fraction of 25\% does not meaningfully change the opacity calculations, and would only marginally affect the dust dynamics due to a lower $\brhod=1.66\,\text{g/cm}^3$.

\section{Results}
\label{sec:results}

In this section we present the results of our numerical simulations. We will first discuss the gap-opening process in the gas, and then explore the effects of dust dynamics on the gap structure.

\subsection{A fiducial model}

We begin by presenting the results of our fiducial model \texttt{rad}, where we assume a coagulation fraction $X_0=0.9$ (i.e., an initial dust-to-gas ratio of $10^{-3}$ in small grains) and ignore the effects of dust feedback in terms of both opacity and backreaction.

\begin{figure}
	\centering
	\includegraphics[width=\columnwidth]{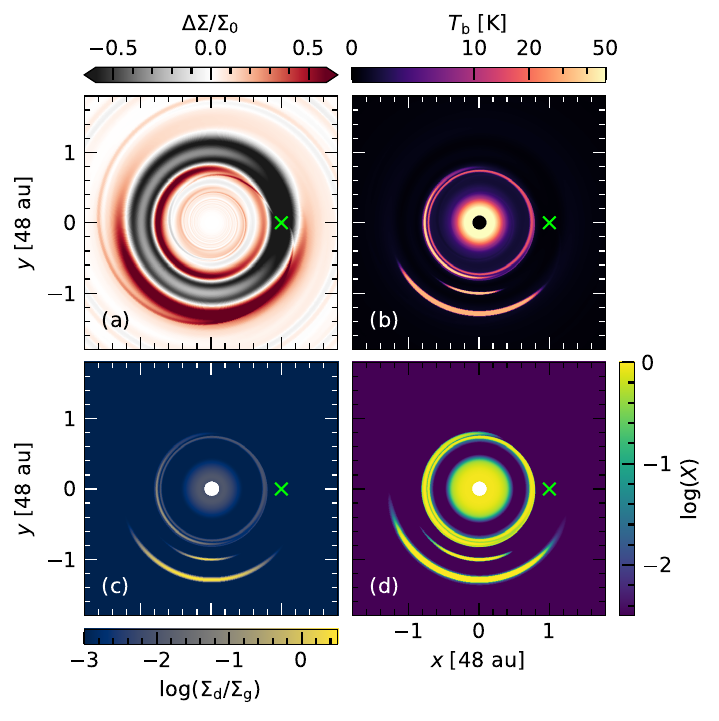}
	\caption{Perturbed gas surface density (panel \emph{a}), dust-to-gas ratio (\emph{b}), dust brightness temperature at 1.25\,mm (\emph{c}) and fraction of dust mass in big grains (\emph{d}) for the fiducial model \texttt{rad} after 1100 planetary orbits. The gas shows gaps, vortices, and pressure bumps, while the dust emission shows bright crescents and rings. A green cross marks the planet's position.}
	\label{fig:fiducial-gas}
\end{figure}

The top left panel of Fig.~\ref{fig:fiducial-gas} shows the perturbed gas surface density after 1100 planetary orbits, where a characteristic set of planet-driven features has been established. Spiral arms permeate the disk, the planet has carved a gap around its orbit, and a large vortex has formed at the gap edge. A secondary, shallower gap is visible in the inner disk, within the pressure bump that has formed at the inner edge of the gap centered of the planet. This inner gap edge shows a slight asymmetry, indicating a weaker vortex that spans a wider azimuthal extent. Finally, a small excess of material can be seen corotating with the planet.

We then plot the dust-to-gas ratio $\varepsilon=\Sigmad/\Sigmag$ in the bottom left panel of the same figure, showing that dust particles have indeed accumulated on the pressure bumps and vortices seen in the gas profile. To create a link to a more readily observable quantity, on the top right panel of Fig.~\ref{fig:fiducial-gas} we show the brightness temperature of the dust emission at a wavelength of 1.25\,mm ($\approx$240\,GHz), computed as
\begin{equation}
	\label{eq:brightness-temperature}
	T_\text{b} = T \left(1-e^{-\tau_\text{d}}\right), \quad\tau_\text{d}=\kappa^\text{big}_\text{1.25\text{mm}}\Sigmad^\text{big} + \kappa^\text{small}_\text{1.25\text{mm}}\Sigmad^\text{small},
\end{equation}
with $\kappa^\text{big}_\text{1.25\text{mm}}=3.1\,\text{cm}^2/\text{g}$, $\kappa^\text{small}_\text{1.25\text{mm}}=2.2\,\text{cm}^2/\text{g}$ the opacity of big and small grains respectively at 1.25\,mm from Fig.~\ref{fig:opacity-wavelength}. As shown on that panel, the mm-sized grains at the outer gap edge have accumulated on the vortex that has formed there, resulting in a bright crescent of emission rather than a ring. At the inner gap edge the dust emission forms a ring instead, with a smaller vortex orbiting next to it. An additional feature due to the material trapped in the planet's corotating region can be seen as well.

We note that the bright inner disk region is simply a temporal artifact, as the mm dust has a smaller $\St$ and therefore drifts inwards slower there. In principle, integrating for a longer time would remove that feature while leaving the rest of the disk structure more or less intact.

All in all, the fiducial model shows the expected features of a planet-driven gap in the gas and the dust emission, with vortices and pressure bumps in the gas profile manifesting as bright crescents and rings in the dust emission. In the following sections we will explore the effects of dust growth and feedback on these structures.

\subsection{Effect of dust growth on gap opening}
\label{sub:dust-growth}

We begin by comparing our fiducial models with different coagulation fractions $X_0$ to explore the effects of dust growth on the gap-opening process. Effectively, by ignoring feedback effects due to gas--dust interaction, varying $X_0$ translates to directly modifying the cooling timescale of the gas, as the dust opacity is reduced (see Fig.~\ref{fig:opacity-temperature}).

The effect of cooling in gap opening has in principle already been explored \citep{miranda-rafikov-2020a,miranda-rafikov-2020b,zhang-zhu-2020,ziampras-etal-2020b,ziampras-etal-2023a}. Nevertheless, considering how weakly constrained the dust fraction of small grains is in protoplanetary disks, the observational significance of incorporating at least a simplified aspect of dust growth in modeling, and how we typically assumed $X\rightarrow0$ in previous work \citep[i.e., a dust-to-gas ratio of 0.01 in small grains, e.g.,][]{ziampras-etal-2020b,ziampras-etal-2023a}, it is useful to comment on this.

\begin{figure}
	\centering
	\includegraphics[width=\columnwidth]{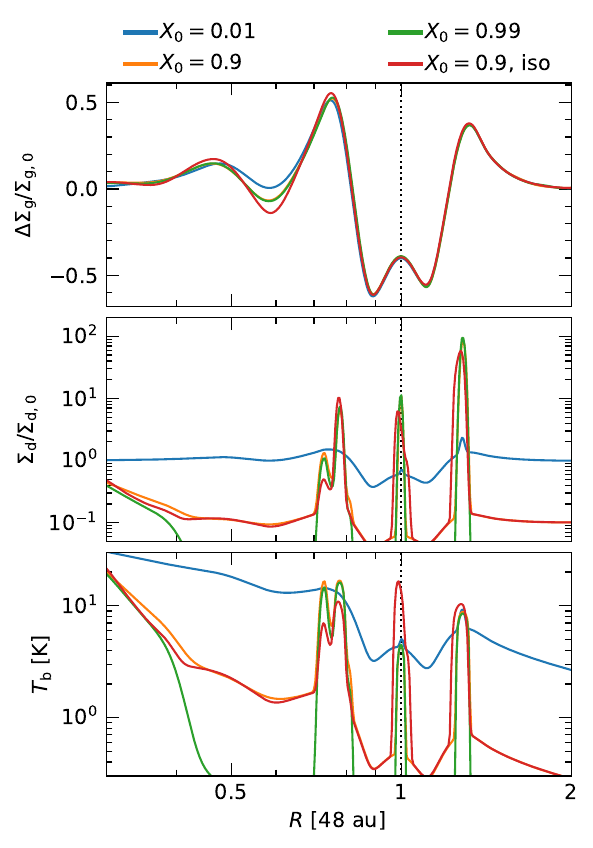}
	\caption{Azimuthally averaged gas surface density (top), dust surface density (middle), and dust brightness temperature at 1.25\,mm (bottom) for models with different coagulation fractions $X_0$. The secondary gap in the gas surface density is shallower for smaller $X_0$ (i.e., more mass in small grains), due to the longer cooling timescale in the inner disk. Substructure in the dust is less pronounced for $X_0=0.01$, due to the well-coupled small grains representing most of the dust mass.}
	\label{fig:compare-X-1D}
\end{figure}
Figure~\ref{fig:compare-X-1D} consolidates our findings for models with $X=0.01$, 0.9, and 0.99, as well as a locally isothermal model with the fiducial $X_0=0.9$. In the top panel we show the azimuthally averaged perturbed gas surface density profile, where the translation of $X$ to a cooling timescale is evident in the depth of the secondary gap at $R\approx0.6\,R_0$. The total dust opacity is significantly higher for $X_0=0.01$ at $T(0.6\,R_0)\sim45$\,K (see Fig.~\ref{fig:opacity-temperature}), which results in a cooling timescale closer to unity and therefore a shallower gap \citep[e.g.,][]{zhang-zhu-2020}. On the other hand, the ``instant'' cooling associated with the locally isothermal prescription leads to a slightly deeper gap. 

It is worth noting that the undulation at $R\approx0.45\,R_0$ does not correspond to a pressure maximum, but rather a local density enhancement with respect to the initial conditions. As a result, mm grains are not trapped at that radius (middle panel of Fig.~\ref{fig:compare-X-1D}) and a corresponding feature is not seen in the dust emission (bottom panel).

The bottom two panels of Fig.~\ref{fig:compare-X-1D} show the azimuthally averaged dust surface density and brightness temperature at 1.25\,mm (240\,GHz) for the same models. For a meaningful mm dust content ($X\gtrsim0.9$) the pressure bump at $R\approx0.75\,R_0$ shows a double peak due to the dust ring and the vortex in the inner gap edge, respectively. The corotating feature and the vortex at the outer gap edge are visible as peaks at $R\approx R_0$ and $R\approx1.3\,R_0$ as well, although they do not correspond to strictly radial structure. It is worth pointing out that, even for $X=0.01$, the submicron grains contribute a nonnegligible amount to the total mm flux due to their comparable opacity at 1.25\,mm. Nevertheless, small bumps are visible for $R\gtrsim R_0$ due to the local accumulation of big grains.

\begin{figure}
	\centering
	\includegraphics[width=\columnwidth]{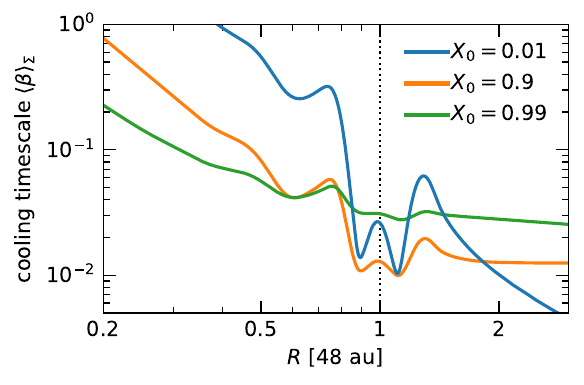}
	\caption{Density-weighted, azimuthally averaged cooling timescale $\beta$ for models with different coagulation fractions $X_0$. While the outer disk and gap region satisfy $\beta\lesssim0.05$ for all $X_0$, the secondary gap at $R\approx 0.6\,\Rp$ has a longer cooling timescale for smaller $X_0$ due to the higher opacity of small grains.}
	\label{fig:beta}
\end{figure}

To illustrate why the differences between models with different $X_0$ are not as pronounced as one might expect in the perturbed gas surface density, we show in Fig.~\ref{fig:beta} the density-weighted, azimuthally averaged cooling timescale $\beta$ computed following \citet{ziampras-etal-2023a}
\begin{equation}
	\label{eq:beta}
	\beta = \frac{1}{f+1}\frac{e}{|\Qcool|}\OmegaK,\qquad f=16\pi \frac{\taueff}{\tauR}\frac{\tauP^2}{6\tauP^2+\pi},
\end{equation}
which results from combining the effects of surface cooling and in-plane radiative diffusion\footnote[1]{We note that \citet{ziampras-etal-2023a} assumed $\tauR=\tauP=\tau$ in their Eq.~(14).}. 
Two takeaways can be drawn from this figure.

First, the cooling timescale has a shallower scaling with radius for higher $X_0$, owing to the lack of a temperature dependence in the opacity for big grains. This results in the profiles for $X_0=0.9$ and $X_0=0.99$ overlapping for $R\sim0.6\,R_0$, and explains why the results between the two models in terms of the location and brightness of substructure in the inner disk are nearly identical.

Second, the cooling timescale at the location of the secondary gap hovers around $\beta\sim0.05$ for $X_0\geq0.9$, with $\beta\sim0.3$ for $X_0=0.01$. This is reflected in the depth of the gap in the gas surface density, and also explains why the gas density profiles agree for all $X_0$ at $R\gtrsim R_0$: in both the primary gap and the outer disk, the cooling timescale is sufficiently short that all models behave similarly in terms of gap opening.

This does not imply that the effects of cooling are negligible, however, as our fiducial radiative model (orange curves in Fig.~\ref{fig:compare-X-1D}) shows differences in the contrast and absolute brightness of radial features in the dust emission compared to a locally isothermal model (red curves in Fig.~\ref{fig:compare-X-1D}). Specifically, the left peak at $R\sim0.75\,R_0$ is significantly weaker in the locally isothermal model, indicating a fainter ring. On the other hand, the corotating region is brighter and slightly wider in the locally isothermal model. This is likely related to the double-trough gap structure found in locally isothermal models \citep[e.g.,][]{cordwell-rafikov-2024} which preserves gas (and therefore traps dust) at $R=\Rp$, as opposed to the slightly more efficient removal of corotating material in radiative models due to planetary spirals dissipating radially closer to the planet \citep{miranda-rafikov-2020a}, which only traps material near the trailing Lagrange point.

\begin{figure}
	\centering
	\includegraphics[width=\columnwidth]{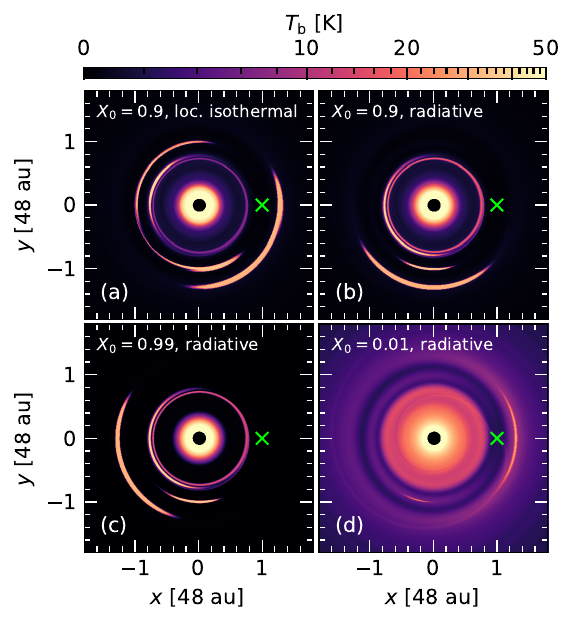}
	\caption{Brightness temperature heatmaps at 1.25\,mm for models with different coagulation fractions $X_0$. The locally isothermal model (panel \emph{a}) shows a fainter ring than its radiative counterpart (\emph{b}) near the inner gap edge and an azimuthally extended bright feature inside the corotating region. The model with $X_0=0.01$ (\emph{d}) is overall brighter and shows less pronounced features, as the bulk of the dust is perfectly coupled to the gas.}
	\label{fig:compare-X}
\end{figure}
Furthermore, we show heatmaps of the dust brightness temperature for all models discussed here in Fig.~\ref{fig:compare-X}. Here, it becomes clear that radial structure is not the full picture. The ring--vortex system observed in the inner disk for our radiative models with $X_0\gtrsim0.9$ nearly fully collapses into a vortex for the locally isothermal model, likely due to the steeper radial density gradient enhancing the Rossby-wave instability \citep{lovelace-1999} and the rapid cooling sustaining the vortex for longer \citep{rometsch-etal-2021,fung-ono-2021}. On the other hand, the model with $X_0=0.01$ shows a rather axisymmetric inner disk, due to the slower cooling resulting in a less pronounced density gradient and a rapid decay of any vortices, as well as a much smaller contribution of big grains to the total mm flux.

As discussed above, the feature in the corotating region of the planet is also more pronounced in the locally isothermal model, spanning over half of the azimuthal extent of the disk, while in the radiative models it is confined to a small region near the Lagrange point behind the planet. Finally, the vortex in the outer gap edge is present in all models, but shows a lower contrast for the model with $X_0=0.01$ due to the substantially higher contribution of small grains to the total mm flux effectively tracing the gas density. This also explains the overall brighter image and less sharp features for $X_0=0.01$.

Overall, while the effect of dust growth is largely predictable when related to the cooling timescale, the impact on the formation of substructures is more nuanced. In our models this manifests as a change in the existence and brightness of features in the dust emission, as well as the presence of different features (vortices vs.~rings) for different models, even though the cooling timescale is $\beta\lesssim0.05$ for the bulk of the disk. 

\subsection{Effect of dust--gas dynamics}
\label{sub:dust-dynamics}

We now turn our attention to the effects of dust--gas interaction on the forming substructures. As discussed in Sect.~\ref{sec:dust-opacity}, the total dust opacity depends on the local fraction of large grains, which evolves dynamically due to the aerodynamic coupling between gas and dust. Our fiducial models have so far ignored this effect, effectively modeling the dust as a static distribution in terms of grain size. In addition, we have not considered the backreaction of the dust on the gas (see Sect.~\ref{sub:dust-model}), which complicates the picture further.

\begin{figure*}
	\centering
	\includegraphics[width=\textwidth]{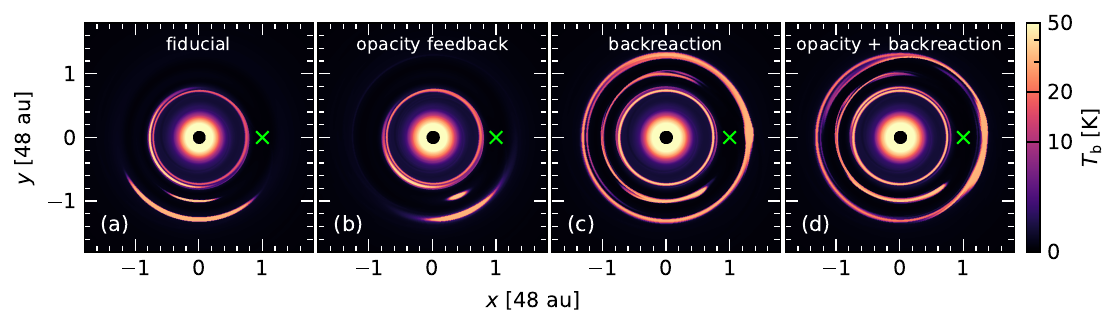}
	\caption{Comparison of the brightness temperature at 1.25\,mm for models with and without opacity feedback and backreaction for the fiducial value $X_0=0.9$ at $t=1100$ planetary orbits. Opacity feedback (panel \emph{b}) results in more compact features, while backreaction (\emph{c}) spreads them out. The combination of the two (\emph{d}) results in a more axisymmetric disk for this value of $X_0$.}
	\label{fig:compare-feedback}
\end{figure*}
Figure~\ref{fig:compare-feedback} shows the results of models with and without opacity feedback and backreaction for the fiducial value $X=0.9$ after 1100~orbits at $R_0$. The effect of opacity feedback is rather subtle, with the vortex at the outer gap edge as well as the corotating feature both being more compact in the azimuthal direction. The features are otherwise similarly bright, with the corotating feature being slightly brighter in the model with feedback. The ring in the inner disk is present in both models, with its brightness practically unchanged.

The effect of backreaction, on the other hand, is clearly visible. Both the vortex at the outer gap edge and the corotating feature are spread out in the azimuthal direction, resulting in an azimuthally extended crescent in the corotating region and a ring in the outer gap edge, with a nonaxisymmetric feature orbiting around it. This effect is significant, as it suggests that turbulent diffusion is not strictly necessary to dissipate vortices into radial features, but rather that a substantial (but realistic) fraction of mass in large grains is sufficient to do so. The ring in the inner disk is also brighter and the local vortex dimmer, but to a lesser extent.

The fact that opacity feedback and dust--gas backreaction have opposite effects, with the former compacting features and the latter spreading them out, suggests that drawing a clear distinction between the two effects in observations may be difficult. Both effects scale with the mass fraction in big grains, and they are likely to be present simultaneously in the outer disk as that region is both cold (i.e., large grains contribute more to the opacity) and rarefied (i.e., mm grains have a larger $\St$ and therefore are less coupled to the gas, see Eq.~\eqref{eq:stokes-number}). Nevertheless, we ran models with both effects for different coagulation fractions $X_0$, as we can compare the differences in the resulting structures directly to Fig.~\ref{fig:compare-X}.

\begin{figure}
	\centering
	\includegraphics[width=\columnwidth]{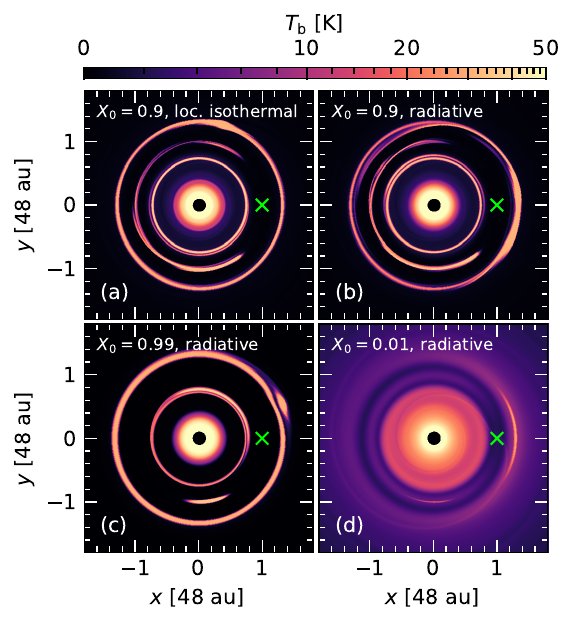}
	\caption{Similar to Fig.~\ref{fig:compare-X}, but for models with opacity feedback and backreaction. For $X_0\geq0.9$, (panel \emph{b}) including these effects results in a ring rather than a vortex at the outer gap edge. For $X_0=0.01$ (\emph{d}) the results are identical to the fiducial model without feedback and backreaction in Fig.~\ref{fig:compare-X}. A small, compact vortex is visible near the outer ring for the locally isothermal (\emph{a}) and $X_0=0.99$ (\emph{c}) models.}
	\label{fig:compare-X-fo}
\end{figure}
Figure~\ref{fig:compare-X-fo} shows the results of models with opacity feedback and backreaction for different coagulation fractions $X_0$. As expected, neither effect is relevant for $X_0=0.01$, where the bulk of the dust is in the perfectly coupled small grains. Interestingly, for $X_0=0.99$, the corotating feature does not smear azimuthally like in the fiducial case with $X_0=0.9$. This suggests that the effect of opacity feedback is more pronounced as $X\rightarrow1$, which makes sense if we consider that the mass fraction in mm grains (which affects backreaction) has increased by $\sim$10\% compared to the case with $X_0=0.9$, but the effective opacity has decreased by $\sim$50\% (compare orange to green curves at $T\sim20$\,K in Fig.~\ref{fig:opacity-approximation}).

In addition, a small vortex-like feature can be seen orbiting just outside of the outer ring for the model with $X_0=0.99$, unlike other models where the radial location of the vortex coincides with the pressure bump at the outer gap edge. This is unlike the classic, massive, azimuthally extended vortices expected from the Rossby-wave instability \citep{lovelace-1999}, as it is instead compact and long-lived. We speculate that this happens in part due to the processes explored by \citet{raettig-etal-2015} and \citet{lovascio-etal-2022}, where the dust--gas backreaction can break large dust-laden vortices into much smaller, more compact features, which also live longer. This is consistent with the fact that such a feature is not present in models without backreaction. We note however that this effect is exaggerated in our 2D framework, as \citet{lyra-etal-2018} have shown that vortices are more resistant to dust--gas-driven dissipation in 3D, being disrupted only near the midplane rather than having the full column of anticyclonic motion destroyed.

\begin{figure}
	\centering
	\includegraphics[width=\columnwidth]{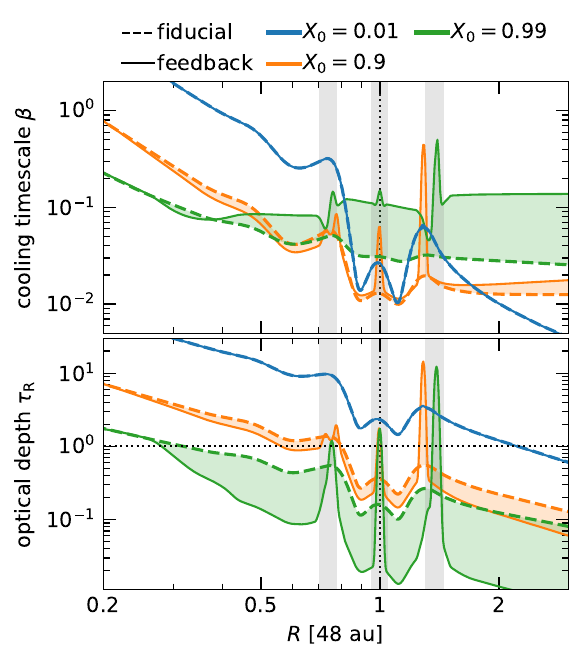}
	\caption{Density-weighted, azimuthally averaged cooling timescale $\beta$ and optical depth $\tauR$ for models with and without opacity feedback. The cooling timescale increases in the optically thin dust-depleted regions due to the lack of opacity-carrying mm grains, and near pressure bumps due to the local enhancement of the dust-to-gas ratio rendering the disk optically thick. The difference between models with and without feedback is filled in to guide the eye. Gray bands approximately mark the location of substructures.}
	\label{fig:tau-beta-feedback}
\end{figure}
To showcase why the effects of dust opacity feedback are more pronounced for higher $X$, we plot in Fig.~\ref{fig:tau-beta-feedback} the cooling timescale $\beta$ and optical depth $\tauR$ for models with and without opacity feedback. Through this figure we can interpret the effects of this mechanism as follows:
\begin{enumerate}
	\item The radial drift of mm grains results in the mass fraction of big grains depleting rapidly in the absence of pressure traps. If the disk is at least moderately optically thick (which is the case for $X_0=0.9$ up to $R\sim R_0$), this results in an overall decrease of the cooling timescale. If the disk is optically thin (outer disk for $X_0=0.9$ and the entire disk for $X_0=0.99$), the cooling timescale is instead increased as the total emissivity of the dust ($\propto\kappaP$) decreases.
	\item Near pressure bumps, the local enhancement of the dust-to-gas ratio results in a significantly higher opacity. This is typically enough to render the disk optically thick, causing the cooling timescale to increase. An exception can be found in the inner ring for $X_0=0.99$, where this local enhancement is just enough to bring the optical depth up to unity. In this case, the cooling timescale is less affected. The high radial opacity gradients due to the accumulation of large grains at pressure bumps also result in very efficient radiative diffusion locally, as discussed in Appendix~\ref{apdx:diffusion}.
\end{enumerate}

Regarding the reason for azimuthal features becoming more compact with opacity feedback, we expect that this happens due to a baroclinic instability \citep{lesur-papaloizou-2010} that is triggered by the local enhancement of the rate of radiative diffusion around dust traps (pressure bumps or vortices) as the flow becomes optically thick (see Fig.~\ref{fig:tau-beta-feedback}) and local temperature gradients are enhanced with respect to the background \citep[see also][]{petersen-etal-2007a,petersen-etal-2007b}. We do note, however, that this argument is speculative in this context, and further investigation and more controlled numerical experiments are needed to understand the underlying mechanism.

We conclude that gas--dust interaction is significant in the formation of substructures in continuum emission, as it can not only alter the azimuthal extent of features but also create entirely new ones. While the effects of opacity feedback and backreaction are opposite, they are both significant in the outer disk ($T\lesssim30$\,K), where mm grains correspond to larger $\St$ and the low temperatures enhance the opacity contribution by larger grains.

\section{Discussion}
\label{sec:discussion}

In this section we discuss the importance of a self-consistent dust model and the implications of our results for the interpretation of continuum observations of protoplanetary disks. We also address the limitations in our work.

\subsection{Towards a self-consistent model of cooling}

Our results show that, even without considering the effects of backreaction on the gas, the redistribution of mm-sized grains around planet-driven substructure can influence the azimuthal extent of features in continuum emission. This is particularly evident in the outer disk, where large grains contribute a nonnegligible fraction of the total opacity.

At the same time, however, in the rarefied, cold outer disk ($R\gtrsim100$\,au), the same large grains would also decouple thermally from the gas, effectively eliminating their contribution to cooling. While we do not consider this effect in our models, it is likely that the cooling timescale in the outer disk is eventually dominated by the small grains, which remain well-coupled to the gas.

In addition, our work assumes that the dust is well-mixed vertically, such that mm-sized grains can provide cooling up to at least a gas scale height. This is not necessarily the case without sufficient vertical mixing, as the mm-sized dust can settle to a thin layer near the midplane such that cooling is effectively mediated solely by the small, vertically-mixed grains. A self-consistent model of cooling would therefore need to consider the height-dependent dust distribution and its footprint in the dust opacity \citep[e.g.,][]{krapp-etal-2024,robinson-etal-2024}, and, ideally, the thermal coupling between gas and dust \citep[e.g.,][]{muley-etal-2023}.

We note, however, that care should be taken to establish that such threedimensional cooling models are accurate in the context of planet-driven gap opening before asserting that they are realistic. For example, \citet{kuiper-etal-2010} have demonstrated that frequency-dependent irradiation yields significantly more accurate midplane temperatures---and therefore cooling timescales---than a gray approximation in optically thin disks. Moreover, \citet{muley-etal-2024} have highlighted the dynamical implications of capturing the shadow cast by the outer gap edge onto the inner disk, which might be suppressed by a diffusive method such as FLD.

Finally, our work assumes a two-population dust model with a fixed coagulation fraction per simulation and a fixed dust composition. Our aim is not to model the full complexity of dust features observed in continuum emission, or even to produce synthetic images of HD~163296, but rather to explore how dust dynamics affect the formation of substructures in the disk. A more detailed model would need to consider a grain size distribution during the coagulation process \citep[e.g.,][]{stammler-birnstiel-2022,robinson-etal-2024} and a more realistic dust composition accounting for the condensation of volatiles \citep[e.g.,][]{semenov-etal-2003}. 

When modeling a specific system, a detailed self-consistent model motivated by observational constraints becomes especially important. In particular, in the context of HD~163296, various studies have constrained the midplane temperature at 48\,au to $\sim22$--25\,K \citep{flaherty-etal-2015,dullemond-etal-2020,law-etal-2021} rather than the 35\,K used in this work. A different temperature (and therefore aspect ratio) would affect gap opening due to the change in the cooling timescale ($\propto h^{-6}$), the spacing between gaps \citep{zhang-etal-2018}, and the formation of vortices \citep{chang-etal-2023}.

\subsection{Turbulent diffusion of dust}

On the note of turbulence, its origin in the outer disk is largely unclear. A popular mechanism that has been proposed as a way to transport dust vertically is the vertical shear instability \citep[VSI,][]{nelson-etal-2013}. However, recent work revealed that the VSI operates more akin to an outwardly propagating wave rather than turbulent eddies \citep{svanberg-etal-2022}. \citet{pfeil-etal-2023} have also shown that dust coagulation can limit the operation of the VSI, due to cooling becoming inefficient as grain growth progresses. For weakly magnetized disks, the ambipolar diffusion-mediated magnetorotational instability can act as a source of turbulence in the outer disk \citep[e.g.,][]{simon-etal-2013}.

Planets can also create meridional flows in their vicinity \citep[e.g.,][]{fung-etal-2019,bi-etal-2021,binkert-etal-2023} or excite vertical motion in the inner disk via the spiral wave instability \citep[SWI,][]{bae-etal-2016a,bae-etal-2016b}, transporting dust vertically. The radial width of dust rings is subject to diffusion due to planet--disk interaction as well \citep{bi-etal-2023}. The efficiency of these processes is disk- and planet-dependent, typically requiring a massive planet. While this can be absorbed to zeroth order as a free $\alpha$-like turbulence parameter in the disk model, the dynamics of dust--gas interaction as a function of height are complex and require dedicated numerical experiments.

In general, regardless of the source of vertical mixing, three-dimensional models are necessary to capture the full dynamics of the dust in the disk in a self-consistent manner. This could be the subject of future work.

\subsection{Observational implications}

In this work we used an HD~163296-lookalike to construct our disk model, but our goal is not to provide a model that can be directly compared to observations of this particular system. Instead, our findings are more general and can be applied to any protoplanetary disk where an embedded planet could be responsible for the formation of substructures in continuum emission.

One implication of our work is that the aerodynamic drag between dust and gas can contribute to at least partially dissolving large-scale vortices near a gap edge into ring-like structures, alleviating the need for significant turbulent diffusion \citep[see also][]{raettig-etal-2015,lyra-etal-2018,lovascio-etal-2022}, provided that a substantial fraction of the dust mass is in large grains. The latter is not an unreasonable requirement, as models of dust growth have shown that dust growth is a rather efficient process \citep[e.g.,][]{birnstiel-2023}. This is particularly relevant for interpreting observed substructures in the context of planet--disk interaction, as addressing the lack of observed vortices is necessary given how common they are in simulations \citep[e.g.,][]{hammer-etal-2017,hammer-etal-2019,hammer-etal-2021,rometsch-etal-2021,fung-ono-2021,chang-etal-2023}. Followup work will discuss the lifetime and observability of vortices with a treatment of radiative cooling and dust--gas dynamics in more detail (Ziampras~\&~Rometsch, in~prep.).

It is also worth highlighting that some of the vortices found in our models might not necessarily be observable as standalone features due to their close proximity to a nearby ring. In an effort to gauge the observability of these features, we present a gallery of simplistic synthetic ALMA observations of our models in Appendix~\ref{apdx:gallery}, after applying a convolution filter with some noise. We find that while the vortices are typically visible, they often appear to merge with the nearby rings, making them difficult to distinguish. Nevertheless, we note the qualitative similarity of the disk structure in our models with full feedback (bottom row of Fig.~\ref{fig:gallery}) to that of HD~143006 \citep[e.g.,][]{andrews-etal-2018}, where a set of two rings and a compact vortex-like feature are observed.

\subsection{Comparison to previous work}
\label{sub:comparison-previous}

Our findings are broadly consistent with previous work both in the context of planet--disk interaction and dust--gas dynamics. Here, we point out some similarities and key differences to previous work.

The efficiency of the gap opening process as a function of the local cooling timescale $\beta$ has been explored by \citet{miranda-rafikov-2020a} and \citet{zhang-zhu-2020}, who found that the formation of multiple gaps is inhibited for $\beta\sim1$. \citet{miranda-rafikov-2020b} then highlighted the importance of in-plane cooling, and \citet{ziampras-etal-2023a} further showed with self-consistent radiative simulations that accounting for in-plane radiative diffusion recovers the aforementioned $\beta\sim1$ criterion (see Eq.~\eqref{eq:beta}). Our results are fully consistent with these findings, highlighted by the shallower secondary gap in the gas surface density for small $X_0$ due to the longer cooling timescale in Sect.~\ref{sub:dust-growth}.

Regarding dust opacity feedback, \citet{krapp-etal-2024} have shown that modeling the dust flow as a standalone fluid and computing the dust opacity self-consistently results in appreciable differences when compared to models that omit this effect in the context of planetary envelopes. While our work is not directly comparable to theirs, as we focus on the formation of substructures in continuum emission rather than the formation of circumplanetary features, we find that the effects of opacity feedback are indeed significant in the planet formation process.

In the context of planet-driven substructure, \citet{binkert-etal-2021} have also performed simulations that included mm-sized dust and both opacity feedback and backreaction. While they did not include a comparison to models without these effects, they recover the formation of two rings about the planetary gap as well as a bright crescent corotating with the planet after $t\approx200$\,orbits at 50\,au for $\Mp=0.3\,\Mjup$ (see Fig.~2 therein). Our results are broadly consistent with theirs, with a key difference being the lack of nonaxisymmetric features in their models due to the very high diffusivity with $\alpha\sim5\times10^{-3}$ that they used.

Finally, regarding the effects of backreaction in the context of planet--disk interaction, \citet{weber-etal-2018} have demonstrated that the resulting gap profile tends to be slightly wider when backreaction is included, due to the inwardly drifting dust driving the local gas outwards via angular momentum exchange. We observe a weaker but similar effect between our fiducial model and the model with backreaction, although we note that the effect is more pronounced in their work due to the higher $\St$ in their models ($\St=0.1$ in Fig.~14 therein as opposed to $\St\sim0.04$ in our models near the gap edge).

\section{Summary}
\label{sec:summary}

In this work we have explored the role of dust--gas dynamics in the context of planet-driven substructures in protoplanetary disks. We carried out a series of numerical simulations of radiation hydrodynamics with dust--gas interaction and an embedded planet, and estimated the resulting dust emission at 1.25\,mm (240\,GHz), corresponding to ALMA Band~6.

To motivate our study, we first constructed a two-population dust model consisting of small (sub-micron) and big (mm-sized) grains, and showed that the total Rosseland and Planck mean dust opacities can be approximated very well with a simple formula that depends on the coagulation fraction $X$ of the small grains. We then showed that the opacity is increasingly dominated by the big grains in the cold, outer disk regions, and that the cooling timescale is therefore sensitive to the local dust size distribution, especially for large $X$.

We first presented the results of our fiducial models where we ignored the effects of dust feedback. We showed that changing $X_0$ effectively adjusted the cooling timescale $\beta$ of the gas by a time-independent factor, resulting in a shallower secondary gap in the gas surface density for small $X_0$. This happened due to the large opacity of the small grains raising the cooling timescale to $\beta\sim0.3$ at $R\approx0.6\,R_0$, which in turn has been shown to inhibit the formation of a secondary gap \citep{miranda-rafikov-2020a,ziampras-etal-2020b,zhang-zhu-2020}. We also showed that the ring in the inner disk was dimmer for an equivalent locally isothermal model, highlighting the importance of radiative cooling even in the quasi-isothermal regime ($\beta\sim0.05$).

We then explored the effects of dust--gas interaction by including opacity feedback, such that the opacity of the dust depended on the local fraction of mm grains, and backreaction, such that the accumulation of mm grains affected the gas dynamics self-consistently. We found that the effects of opacity feedback and backreaction were opposite, with the former compacting features such as vortices in azimuth and the latter instead spreading them out into ring-like structures. For our fiducial case of $X_0=0.9$, the combination of both effects resulted in a nearly axisymmetric disk, as the vortex at the outer gap edge and the dust accumulation corotating with the planet were both smeared out azimuthally. The two effects were more pronounced for larger $X_0$, and appeared to compete for $X_0=0.99$, where the corotating feature remained compact and a small vortex was still visible near the outer ring.

By analyzing the cooling timescale and optical depth profiles of models with and without opacity feedback as a function of $X_0$, we showed that the trapping of mm grains near pressure bumps can render the disk optically thick even if 90--99\% of small grains have coagulated into big grains. This in turn resulted in a significant increase in the cooling timescale locally, stressing the highly nonlinear nature of dust--gas interaction in the outer disk.

Our results suggest that the aerodynamic drag between dust and gas can dissolve large-scale vortices near pressure bumps, with the process being quite efficient for reasonable dust-to-gas ratios in big grains. This has implications for the interpretation of observed substructures in protoplanetary disks, as the rarity of observed nonaxisymmetric features may be due to the effects of dust--gas interaction rather than turbulent diffusion. This could help relax the tension between theory and observations in light of recent work on hydrodynamical instabilities. Followup work will investigate in more detail the conditions under which planet-driven vortices can be observed in continuum emission (Ziampras~\&~Rometsch, in~prep.).

\section*{Acknowledgments}
This research utilized Queen Mary's Apocrita HPC facility, supported by QMUL Research-IT (http://doi.org/10.5281/zenodo.438045). This work was performed using the DiRAC Data Intensive service at Leicester, operated by the University of Leicester IT Services, which forms part of the STFC DiRAC HPC Facility (www.dirac.ac.uk). The equipment was funded by BEIS capital funding via STFC capital grants ST/K000373/1 and ST/R002363/1 and STFC DiRAC Operations grant ST/R001014/1. DiRAC is part of the National e-Infrastructure. Computations were also performed on the \texttt{astro-nodes} at MPIA. AZ and RPN are supported by STFC grant ST/P000592/1, and RPN is supported by the Leverhulme Trust through grant RPG-2018-418. PS acknowledges the support of the DFG through grant number 495235860 and is a Fellow of the International Max Planck Research School for Astronomy and Cosmic Physics at the University of Heidelberg (IMPRS-HD). All plots in this paper were made with the Python library \texttt{matplotlib} \citep{hunter-2007}. Typesetting was expedited with the use of GitHub Copilot, but without the use of any AI-generated text.

\section*{Data Availability}

Data from our numerical models are available upon reasonable request to the corresponding author.

\bibliographystyle{mnras}
\bibliography{refs}


\appendix

\section{Dust dynamics in \pluto{}}
\label{apdx:dust-pluto}

In this section, we document the numerical implementation of the dust--gas coupling term in the \pluto{} code. We briefly describe the algorithm for a two-fluid module for dust and gas as it is available in the \pluto{} \texttt{v4.4} astrophysical code, and present a series of tests to validate our implementation.

Dust--gas interaction is key to understanding the initial conditions for planet formation, as scarce dust is what forms planets. Dust in numerical codes is largely modeled as Lagrangian particles \citep[e.g.,][]{bai-stone-2010,yang-johansen-2016,price-etal-2018,mignone-2019}, but in regimes where dust is sufficiently coupled to the gas ($\St\lesssim 1$), or high particle concentration, one can model it as a pressureless fluid. This was first implemented in astrophysical fluid dynamics codes such as \texttt{PIERNIK} \citep{kowalik-etal-2013}, \texttt{MPI-AMRVAC} \citep{porth-etal-2014}, \texttt{LA-COMPASS} \citep{fu-etal-2014}, and also more recently in \texttt{FARGO3D} \citep{benitez-etal-2019} and \texttt{ATHENA++} \citep{huang-bai-2022}. These codes that employ multiple dust species prefer implicit or semi-implicit solvers for the dust--gas interaction to avoid stiff regimes for large collision rates \citep{stoyanovskaya-etal-2018}.

We note that the explicit solver described and tested in this section is already available as of version \texttt{4.4} of the code, whereas the implicit solver will become available in a subsequent update. Support for additional dust species as well as a second order implicit-explicit (IMEX) scheme is currently under development, and will be released in the future.

\subsection{Numerical implementation}
\label{sub:dust-implementation}

We describe the algorithm and tests for a two-fluid module for dust and gas implemented in the \pluto{}~\texttt{v4.4} hydrodynamics code. Dust interacts with gas via aerodynamic drag following dynamics outlined by the continuity and momentum equations (see Eqs.~\eqref{eq:navier-stokes-1}~\&~\eqref{eq:navier-stokes-2} in a vertically integrated framework). Following the prescription in \citet{benitez-etal-2019} for an arbitrary number of species, we write down the equations for two species (i.e., gas or dust) where index $i \in [\text{g}, \text{d}]$, 
\begin{subequations}
	\label{eq:3ddustfluid}
	\begin{align}
		\label{eq:continuity-equations}
		&\DP{\rho_\text{i}}{t} +  \nabla \cdot (\rho_i\vel_i) = 0\\ &\DP{~(\rho_\text{i}\vel_i)}{t} + \nabla \cdot(\rho_\text{i} \vel_i \otimes \vel_i + \delta_{i\mathrm{g}}P_i \mathbb{I} ) =  f_i,
	\end{align}
\end{subequations}
with the identity matrix $\mathbb{I}$, and the drag force defined in terms of the collision rate $\alpha_{ij}$
\begin{equation}
	\label{eq:dust-dragterm}
	f_i = - \rho_i \alpha_{ij} (\vel_i - \vel_j), \; i \neq j 
 \end{equation}
where $\alpha_{ij} = a_i \delta_{j\text{g}} + \epsilon_j a_j \delta_{\text{g}i}$, with $\delta$ being the Kronecker delta function. The collision rate is parameterized using the stopping time $\ts$, giving $a_{i} = \ts^{-1}$.
We only include the contribution of dust in momentum transfer and hence solely focus on conservation of momentum, and not total energy. Dust diffusion is currently not part of the module.

\begin{figure}
	\centering
	\includegraphics[width=\columnwidth]{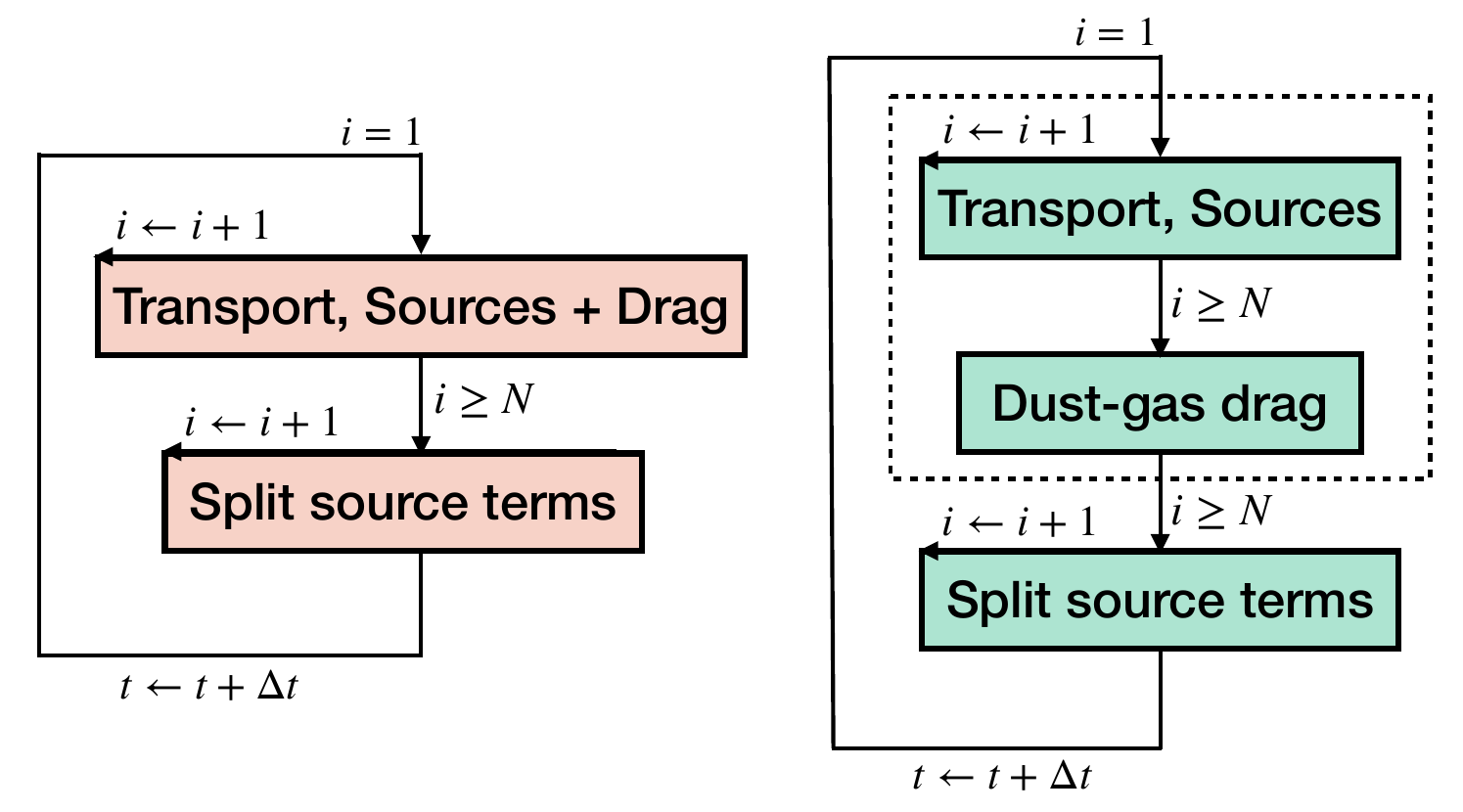}
	\caption{A simplified flowchart of the steps involved in the explicit (left) and implicit (right) schemes including dust drag. The drag term is added to the transport and source term stages in the case of the explicit solver, while for the implicit solver, operator splitting is used. The dotted box is used to indicate that transport, source terms, and dust--gas coupling are treated as a single, ``unsplit'' step for the purposes of Strang splitting in order to recover the terminal velocity limit for the dust \citep{booth-etal-2015}.}
	\label{fig:flowchart}
\end{figure}

The way transport and source terms are solved for in the \pluto{} code (see Fig.~1 in \citet{mignone-etal-2007} for the flowchart) can be described using the reconstruct-solve-average (RSA) method:
\begin{enumerate}
\item \noindent volume-averaged conserved quantities $\left(\rho,\rho\vel, e+\rho u^2/2\right)$ are converted to their primitive counterparts $\left(\rho, \vel, e\right)$,
\item \noindent left- and right-interface values are reconstructed from volume averages by means of piecewise polynomial interpolands in each zone,
\item \noindent a Riemann problem is solved to calculate numerical fluxes for the gas
\item \noindent fluxes are used to update the conserved quantities keeping the whole scheme unsplit and explicit, and the solution is advanced with a timestep determined by the Courant--Friedrichs--Levy condition (CFL).
\end{enumerate}
\pluto{} employs Strang splitting to handle source terms where physical processes, such as viscosity and cooling, operate on different timescales compared to transport. We integrate our dust--drag terms into this framework as described by the flowchart in Fig.~\ref{fig:flowchart}. For the dust, numerical fluxes are determined by the Riemann solver in \citet{leveque-2004}.

The public version of the dust module uses an explicit integrator, including Eq.~\eqref{eq:dust-dragterm} as an additional source term and thereby achieving second-order temporal accuracy. 

We also implemented an unconditionally stable implicit scheme that is used in the article (see Section~\ref{sub:numerics}). The latter updates the velocities of both fluid components with an operator-split step that is exact to first order in time, following equations \eqref{eq:dust-evolution-2} and \eqref{eq:back-reaction}:

\begin{equation}
\label{eq:dust-update}
  \left.
    \begin{aligned}
      & \rho_\text{g}\frac{\velg' - \velg}{\Delta t} = -\rho_\text{d}\frac{\velg'-\veld'}{\St}\OmegaK \\
      & \rho_\text{d}\frac{\veld' - \veld}{\Delta t} = -\rho_\text{d}\frac{\veld'-\velg'}{\St}\OmegaK
    \end{aligned}
  \right\} \Rightarrow
    \begin{aligned}
      & \velg' = \frac{(1+b)\velg + b\epsilon\veld}{1+b+b\epsilon}\\
      & \veld' = \frac{b\velg + (1+b\epsilon)\veld}{1+b+b\epsilon}
    \end{aligned}
\end{equation}
%
where $\vel'$ indicates the velocity after a timestep $\Delta t$, $b \equiv \Delta t/\ts$ is the ratio of timestep to stopping time, and $\epsilon \equiv \rho_\text{d} / \rho_\text{g}$ is the dust-to-gas ratio.

To ensure that the correct terminal velocity limit is recovered, the dust--gas drag operator in Eq.~\eqref{eq:dust-update} is always applied after the transport and source terms regardless of whether additional, split-source terms are considered \citep[][see also Fig.~\ref{fig:flowchart}]{booth-etal-2015}. Although it reduces the accuracy of the scheme to first-order, benefits of employing an unconditionally stable, fully implicit integrator are compared to those of a fully explicit one in Sect.~\ref{sub:dust-convergence}.
\subsection{Testing}
In this section, we validate our solver using some well-known tests with checks for stability, accuracy and convergence. Unless otherwise stated, tests have been performed using the HLLC Riemann solver \citep{toro-etal-1994}, linear spatial reconstruction, the limiter by \citet{vanleer-1974}, and a second-order Runge--Kutta time integrator.

\subsubsection{Radial drift test}
\label{sub:dust-radial-drift}

An unperturbed disk rotates at sub-Keplerian velocities due to the background pressure gradient. Dust does not feel pressure and therefore typically feels a head-/tailwind, drifting in-/outwards. For a simple accreting disk in the absence of dust feedback the dust velocity can be written analytically \citep{takeuchi-lin-2002} as
\begin{equation}\
	\label{eq:radial-drift}
    u_{R,\text{d}} = \frac{\eta R\OmegaK + \St^{-1} u_{R,\text{g}}}{\St + \St^{-1}}, \qquad \eta = \D{\log P}{\log R} h^2, \quad P = \Sigma \cs^2,
\end{equation}
where $\St = \ts \OmegaK$ is the Stokes number.

We set up a 1D vertically integrated, inviscid, adiabatic disk in polar coordinates, with 1024 logarithmically-spaced radial cells and $R\in[0.4, 2.5]~R_0$. As initial conditions, we use power laws for the surface density $\Sigma = \Sigma_0(R/R_0)^{-1/2}$, temperature $T = T_0(R/R_0)^{-1}$, and a constant aspect ratio $h = 0.05$. 
We set the radial boundaries to the initial profile, and use a linear damping function at the edges of the domain. We switch off dust feedback and run the model for different $\St \in [10^{-3}, 100]$ in intervals of powers of 10. We then integrate for $100\,\Omega_K^{-1}$ to ensure that the disk has reached a steady state, with a CFL of 0.4. Under these conditions, Eq.~\eqref{eq:radial-drift} simplifies to
\begin{equation}
	\label{eq:radial-drift-simple}
	\frac{u_{R,\text{d}}}{\vK} = -\frac{3}{2}\frac{h^2}{\St + \St^{-1}} = \text{const.}, \quad \vK = R\OmegaK.
\end{equation}
We therefore compute the radial average of $u_{R,\text{d}}/\vK$ between 1--2\,$R_0$ and compare it to the theoretical expectation from Eq.~\eqref{eq:radial-drift-simple}.

\begin{figure}
	\centering
	\includegraphics[width=\columnwidth]{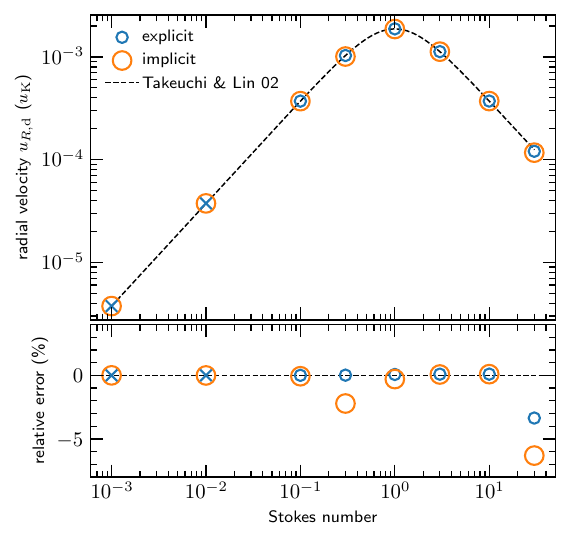}
	\caption{Comparison of radial velocity values of the dust component between the explicit \pluto{} dust fluid module (release version \texttt{4.4}, blue) and the implicit module (used in this paper, orange) against analytical profiles (dashed lines) from \citet{takeuchi-lin-2002} for different Stokes numbers, measured by averaging $u_{R,\text{d}}/\vK$ between 1--2\,$R_0$. For $\St \lesssim 0.001$ (indicated by crosses), the explicit method is only stable for smaller time steps (we use a CFL = 0.2 for the $\St = 10^{-2}$ model, and a fixed time step of $\Delta t = 10^{-5}$ for the $\St = 10^{-3}$ model). The bottom panel shows the relative errors between each result and its theoretical value. The explicit solver performs better overall, while both models show higher errors for $\St > 10$.}
	\label{fig:vdrift-plot}
\end{figure}

We plot the results of the comparison after $100\,\Omega_K^{-1}$ in Fig~\ref{fig:vdrift-plot}. Both the implicit and explicit solvers match with the analytical profile very well, with relative errors within 2\%, except for $\St > 20$ which is reasonable as the fluid approximation breaks down for weak dust--gas coupling. The errors are smaller for the second-order explicit solver, but we note that for $\St < 0.01$, the explicit solver needs significantly smaller time steps to maintain stability, in contrast to the unconditionally stable implicit scheme. We conclude that our solvers perform well without dust feedback.

\subsubsection{Linear $\dw{}$ test}
\label{sub:linear-dw-test}

\begin{figure}
	\centering
	\includegraphics[width=\columnwidth]{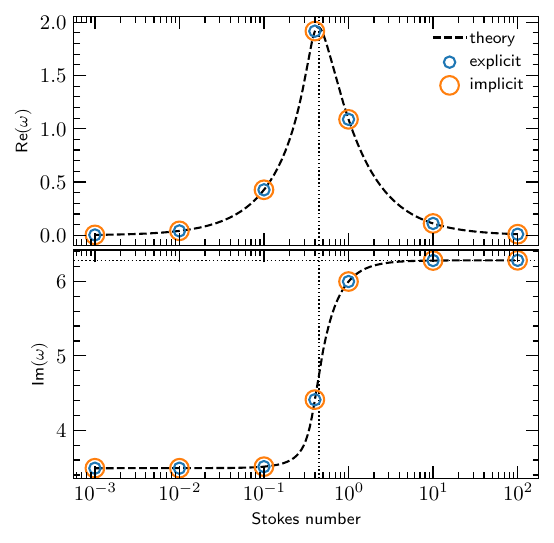}
	\caption{Real and imaginary components of the linear dusty wave for the explicit and implicit dust fluid modules. The dashed curves correspond to the analytical profiles, and a vertical dotted line corresponds to the value of $\St$ for which the wave damps the fastest. The eigenvectors for the different Stokes numbers can be found in Table~\ref{tab:eigenvalues}.}
	\label{fig:dg-coupling}
\end{figure}

We test dust--gas coupling in our solvers using the $\dw{}$ test \citep{laibe-price-2011, laibe-price-2012} that involves adding a perturbation to the two-fluid system and comparing its oscillation and damping rate with expectations from linear theory. Similar to the prescription in equations~(45)~\&~(46) of \citet{benitez-etal-2019}, for two species, a perturbation of the form $\delta p = \delta \hat{p} e^{ikx - wt}$ where $k$ is a wavenumber gives the dispersion relation
\begin{equation}
    \omega^3 \ts - \omega^2 (1 + \epsilon) + \omega \omega_\text{s}^2 \ts - \omega_\text{s}^2 = 0
\end{equation}
where $\omega_\text{s} = k \cs$. This is solved for different values of the stopping time $\ts$ (which, in dimensionless units, represents the Stokes number $\St$) and the complex roots which describe the oscillation and damping rate of the sound wave are calculated. For $\omega \neq \ts^{-1}$, the components of the eigenvectors for all quantities are given by
\begin{subequations}
	\label{eq:eigenvectors}
	\begin{align}
		\label{eq:velg}
  \frac{\delta \hat{u}_\text{g}}{\cs} = - \frac{\omega}{\omega_\text{s}} \frac{\delta \hat{\rho}_\text{g}}{\rho_\text{g}^0}
         \end{align}
         \begin{align}
		\label{eq:veld}
  \frac{\delta \hat{u}_\text{d}}{\cs} = - i \frac{\omega}{\omega_\text{s}} \frac{1}{(1 - \omega \ts)} \frac{\delta \hat{\rho}_\text{g}}{\rho_\text{g}^0} \; \text{and}
         \end{align}
         \begin{align}
		\label{eq:rhod}
  \frac{\delta \hat{\rho}_\text{d}}{\rho_\text{d}^0} =  \frac{1}{(1 - \omega \ts)} \frac{\delta \hat{\rho}_\text{g}}{\rho_\text{g}^0} .  \end{align}
\end{subequations}
For different Stokes numbers, we list the different eigenvectors for the dust density, gas and dust velocities in Table~\ref{tab:eigenvalues}.
We calculate numerical solutions for this problem with a setup with a constant background density $\rho_\text{d}^0 = 2.24$, $\rho_\text{g}^0 = 1$, and zero velocities, adding a perturbation of the form,
\begin{equation}
    \label{eq:perturbation}
    \delta p = A~[\text{Re}(\delta\hat{p})~\text{cos}(kx) -\text{Im}(\delta\hat{p})~\text{sin}(kx)].
\end{equation}
where a small amplitude $A = 10^{-4}$ is used. We use periodic boundary conditions, a domain of size $x \in [0, L]$ with $L = 1$ and 1024 cells, and set the sound speed $\cs = 1$ and the wavenumber $k=2\pi/L$. We then compare the damping rate Re$(\omega)$ and oscillation frequency Im$(\omega)$ to the analytical solution and find very good agreement with our implicit and explicit solvers over the range of Stokes numbers as illustrated in Fig.~\ref{fig:dg-coupling}. The second-order explicit scheme shows lower errors overall, with relative errors below 0.1\%. The implicit scheme shows larger errors in the damping rate $\text{Re}(\omega)$ for the low Stokes regime, with a highest relative error of 10\% for $\St = 10^{-3}$. 

\begin{table*}
  \centering
  \begin{tabular}{ l  c  c  c }
    \hline
    Stokes No. ($\St$) & $\delta \hat{\rho}_\text{d}/\rho_\text{d}^0$ & $\delta \hat{u}_\text{g}/\cs$ & $\delta \hat{u}_\text{d}/\cs$ \\ 
    \hline

    0.001 &  $0.75902381 + 1.06158633i$ & $0.55555587 - 0.00067036i$ & $0.55555378 + 0.00126889i$ \\
    0.01 & $0.75513887 + 1.07250525i$ & $0.55558735 - 0.00670456i$ & $0.55537838 + 0.01268826i$\\
    0.1 & $0.69802164 + 1.18268581i$ & $0.55890819 - 0.06801951i$ & $0.53758089 + 0.12615568i$ \\
    0.4 & $0.16525076 +  1.24780066i $ & $0.70195945 - 0.30492432i$ & $0.22164469 + 0.36853415i$ \\
    1 & $0.20811519 + 0.88463982i$ & $0.95465712 - 0.17330030i$ & $0.02652757 + 0.15954801i$ \\
    10 & $0.29621481 + 0.77903121i$ & $0.99955732
    - 0.01782083i$ & $0.00025341 + 0.01591598i$ \\
    100 & $0.30515898 + 0.77039615i$ & $0.99999557
    - 0.00178253i$ & $0.00000025 + 0.00159155i$ \\
    \hline
  \end{tabular}
  \caption{Initial conditions for the \dw{} test, corresponding to eigenvectors for perturbations in the dust density, gas and dust velocities for $\St \in [10^{-3}, 100]$. Perturbations in gas density are always assumed to be $\delta \hat{\rho}_\text{g}/\rho_\text{g}^0 = 1$. These values are used in Eq.~\eqref{eq:perturbation}.}
  \label{tab:eigenvalues}
\end{table*}
\subsubsection{Convergence}
\label{sub:dust-convergence}

To evaluate the spatial convergence of our solvers, we repeat the \dw{} tests in Sect.~\ref{sub:linear-dw-test} for different grid resolutions. To achieve this, we vary the number of grid cells in the range $N_x \in [128, 2048]$ in intervals of powers of 2. Similarly to \citet{benitez-etal-2019}, we allow the time-step to vary by choosing a CFL of 0.4.
We then calculate the root mean square error (RMSE) for any quantity, here $\rho_\text{g}$, defined by
\begin{equation}
    \label{eq:MSE}
    \text{RMSE} = \sqrt{\frac{1}{N} \sum\limits_\text{cells} \left(\rho_\text{g} - \rho_\text{g}^{\text{theory}}\right)^2}
\end{equation}

\begin{figure}
	\centering
	\includegraphics[width=\columnwidth]{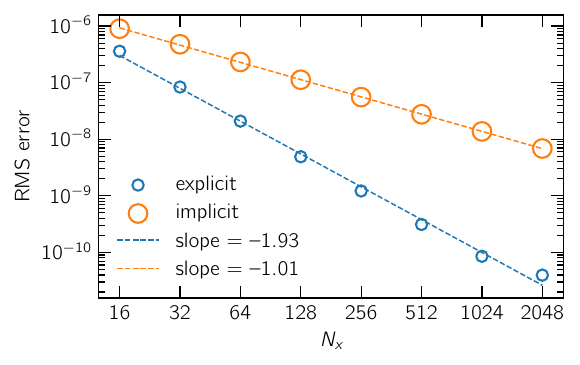}
	\caption{Root mean square errors for the dusty wave test models with different spatial resolutions. The implicit module shows an error that scales linearly with $N_x$, while the fully second-order explicit module scales quadratically.}
	\label{fig:space-plot}
\end{figure}

We see very good convergence for both the implicit and explicit solvers, with the overall errors being lower for the second-order explicit solver as seen in Fig.~\ref{fig:space-plot}
The scaling of the RMSE with spatial resolution agrees very well with that of Fig.~6 in \citet{benitez-etal-2019}, which shows a mean square error scaling with $N_x^{-2.2}$ (i.e., $\text{RMSE}\propto N_x^{1.1}$). We therefore conclude that both the explicit and implicit dust fluid modules implemented in the \pluto{} code are well-tested in terms of stability, accuracy, and convergence rate.

\section{Effect of turbulent diffusion}
\label{apdx:alpha}

In the main text we have focused on dust--gas interaction while ignoring the diffusive effects of turbulence, which has nevertheless been shown to be important in the context of planet-driven substructures \citep[e.g.,][]{zhang-etal-2018}. In the interest of comparing with previous work while also examining the effects of diffusion within the scope of our study, we have performed a set of simulations based on our model with opacity feedback and backreaction for our fiducial value of $X_0=0.9$ and with varying $\alpha\in\{10^{-5}, 10^{-4}, 10^{-3}\}$.

\begin{figure}
	\centering
	\includegraphics[width=\columnwidth]{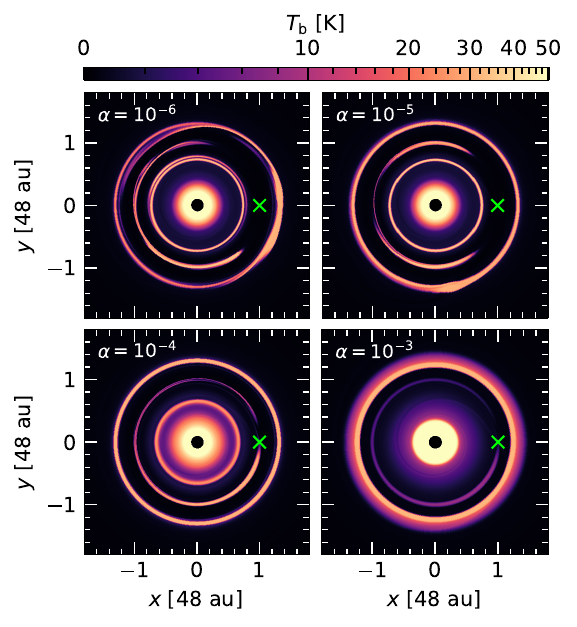}
	\caption{Similar to Fig.~\ref{fig:compare-X-fo}, but varying the $\alpha$ parameter while including opacity feedback and backreaction for $X_0=0.9$.}
	\label{fig:compare-alpha}
\end{figure}

The results at $t=1100$ planetary orbits are shown in Fig.~\ref{fig:compare-alpha}, where we compare the brightness temperature at 1.25\,mm for the different values of $\alpha$ used. We find that a low value of $\alpha=10^{-5}$ is sufficient to dissolve the vortex orbiting about the inner ring at $R\approx0.75$\,au, and for a moderate $\alpha=10^{-4}$ the vortex at the outer gap edge has dissipated as well. For an even higher value of $\alpha=10^{-3}$, viscous diffusion is strong enough to prevent the carving of the secondary gap at $R\approx0.6\,R_0$ and the formation of the inner ring, in agreement with the findings of \citet{zhang-etal-2018} (see Fig.~6~\&~7 therein for $h=0.05$--0.07).

\section{Radiative diffusion around rings}
\label{apdx:diffusion}

\begin{figure}
	\centering
	\includegraphics[width=\columnwidth]{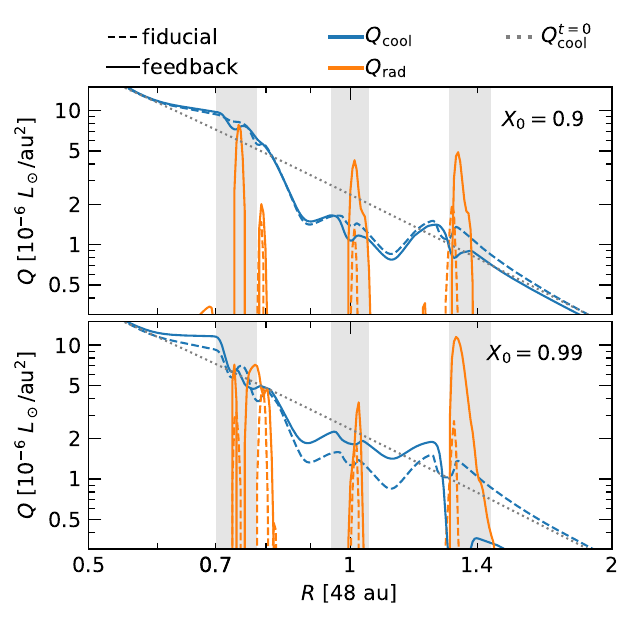}
	\caption{Radial profiles of the cooling terms in Eq.~\eqref{eq:source-terms-3}~\&~\eqref{eq:source-terms-4} at $t=1100$ planetary orbits for the fiducial model and that with opacity feedback and backreaction for $X_0=0.9$ and 0.99. While in-plane cooling ($\Qrad$) is typically orders of magnitude weaker than vertical cooling ($\Qcool$), it dominates around the rings due to the significant optical depth gradients induced by the accumulation of dust grains. Gray bands approximately mark the location of substructures.}
	\label{fig:source-terms}
\end{figure}

In Sect.~\ref{sub:dust-dynamics} we discussed how the accumulation of big grains at pressure bumps can increase the opacity in a radially narrow region. While this translates to a change in the cooling timescale as seen in Fig.~\ref{fig:tau-beta-feedback}, the radial gradients of $\kappaP$ and $\kappaR$ can also lead to much more efficient in-plane cooling through radiative diffusion in the vicinity of a ring. To showcase this effect, we plot radial profiles of the terms $\Qcool$ and $\Qrad$ in Eqs~\eqref{eq:source-terms-3}~\&~\eqref{eq:source-terms-4} at $t=1100$\,orbits for our fiducial model and that with opacity feedback and backreaction for $X_0=0.9$ and 0.99 in Fig.~\ref{fig:source-terms}.

Here, it becomes clear that in-plane cooling becomes important around the rings, highlighting the importance of radiative diffusion in the context of planet-driven substructures. At the same time, the opacity feedback from the pileup of big grains further increases the efficiency of in-plane cooling, and especially so for larger $X_0$. In particular, $\Qrad$ increases by a factor of 2.5 and 5 in the outer ring for $X_0=0.9$ and 0.99, respectively, compared to the fiducial model without feedback.

\section{Gallery of mock ALMA observations}
\label{apdx:gallery}

Here we present a gallery of the brightness temperature $\Tb$ at 1.25\,mm models used in our study at a timestamp of 1100 planetary orbits ($\sim$0.26\,Myr) and after applying a filter to account for typical observational limitations in a simplistic manner. To do this, we first add random noise with an amplitude of up to 2\,K to $\Tb$ and then convolve the result with a Gaussian beam with a radius of 4\,au. The latter corresponds to $40$\,mas assuming a disk at a distance of 100\,pc.

\begin{figure}
	\centering
	\includegraphics[width=\textwidth]{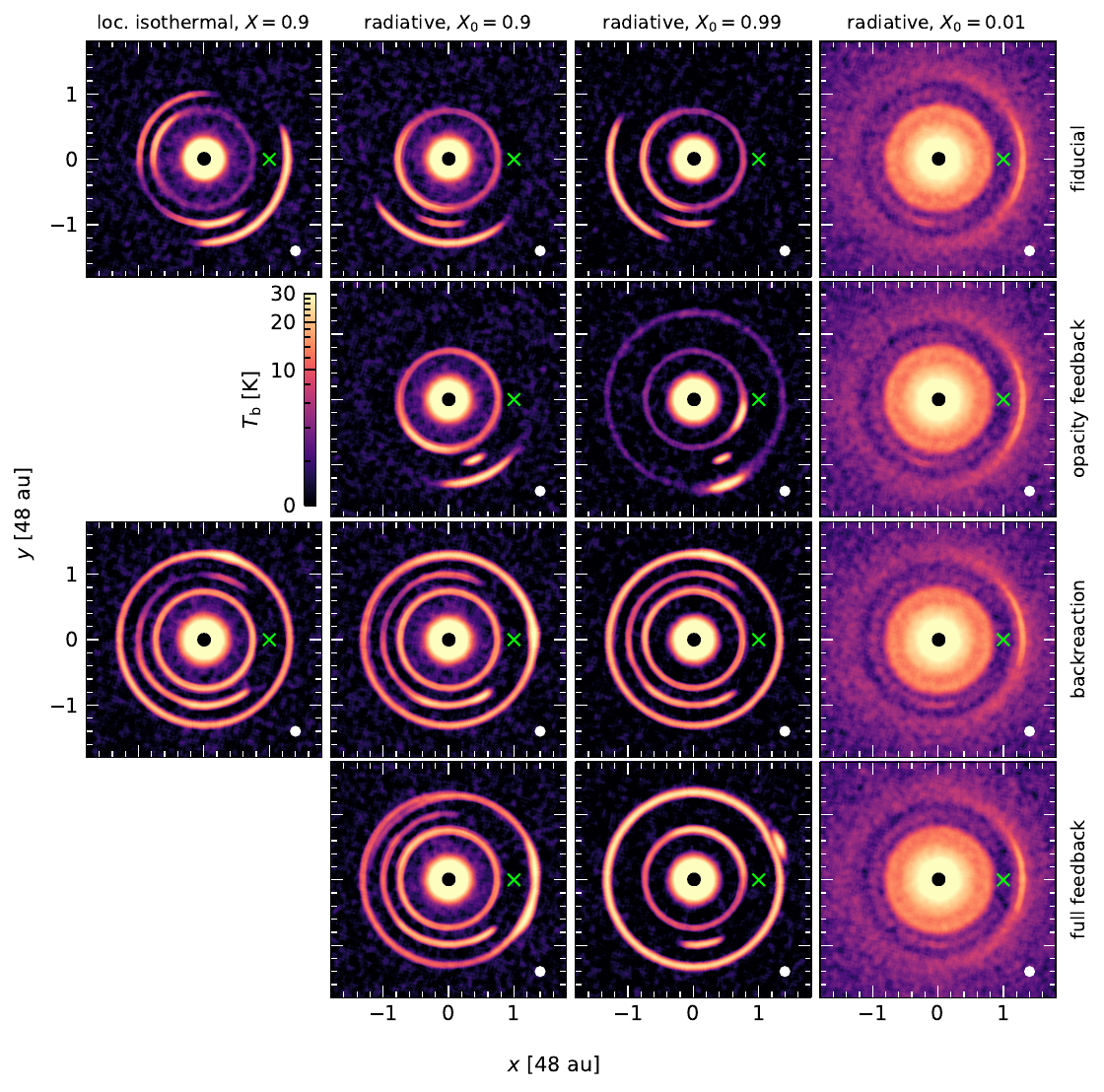}
	\parbox{\textwidth}{\caption{Brightness temperature heatmaps at 1.25\,mm for all models at $t=1100$~planetary orbits, after adding random noise and applying a convolution filter with a beam size of 4\,au (see white dots on the bottom right of each panel).}}
	\label{fig:gallery}
\end{figure}

The results are shown in Fig.~\ref{fig:gallery}. The vortices in the inner disk, while clearly separated from their neighboring rings before convolving, now typically appear to merge with them due to the beam size. The addition of noise makes it difficult to distinguish radial features when they are faint (e.g., the inner ring in the fiducial locally isothermal model and both rings in the model with $X=0.99$ and opacity feedback), as well as the corotating dust bumps for $X=0.01$. 


\bsp	
\label{lastpage}
\end{document}